%% file: main_camera_ready.tex
\documentclass[conference]{IEEEtran-4}

\usepackage{tikz}
\usepackage{amsmath}
\usepackage{xcolor,soul}
\usepackage{tcolorbox}
\tcbuselibrary{breakable}
 \usepackage{enumitem}
 \usepackage{amsfonts}
\usepackage{xcolor}
\usepackage{tcolorbox}
\usepackage{algorithm} 
\usepackage{algpseudocode}

 \usepackage{multirow}
\usepackage{graphicx}
\usepackage{subcaption}
\usepackage{rotating}
\usepackage{adjustbox}
\usepackage{mwe}
\usepackage[htt]{hyphenat}
\usepackage{setspace}
\usepackage{float}
\usepackage{placeins}
\usepackage{url}
\usepackage{listings}

\usepackage{pifont}     
\usepackage{multirow}   
\usepackage{arydshln}    

\newcommand{\cmark}{\ding{51}} 
\newcommand{\xmark}{\ding{55}} 

\usepackage{booktabs}
\usepackage{colortbl}
\usepackage[normalem]{ulem}
\useunder{\uline}{\ul}{}
\usepackage[colorlinks=false,pdfborder={0 0 0}]{hyperref}

\ifCLASSINFOpdf
\else
\fi
\begin{document}
%
\title{Memory Backdoor Attacks on Neural Networks}




%
\author{\IEEEauthorblockN{Eden Luzon\IEEEauthorrefmark{1}, Guy Amit\IEEEauthorrefmark{1}, Roy Weiss\IEEEauthorrefmark{1}, Torsten Krauß\IEEEauthorrefmark{2}, Alexandra Dmitrienko\IEEEauthorrefmark{2} and Yisroel Mirsky\IEEEauthorrefmark{1}\IEEEauthorrefmark{3}\thanks{\IEEEauthorrefmark{3}Corresponding Author.}}
\IEEEauthorblockA{\IEEEauthorrefmark{1}Ben-Gurion University, Institute of Software Systems and Security. \IEEEauthorrefmark{2}University of Würzburg}
\IEEEauthorblockA{\{luzone, guy5, weissroy\}@post.bgu.ac.il, \{torsten.krauss, alexandra.dmitrienko\}@uni-wuerzburg.de, yisroel@bgu.ac.il}
}


\IEEEoverridecommandlockouts
\makeatletter\def\@IEEEpubidpullup{6.5\baselineskip}\makeatother
\IEEEpubid{\parbox{\columnwidth}{
		Network and Distributed System Security (NDSS) Symposium 2026\\
		23 - 27 February 2026 , San Diego, CA, USA\\
		ISBN 979-8-9919276-8-0\\  
		https://dx.doi.org/10.14722/ndss.2026.241870\\
		www.ndss-symposium.org
}
\hspace{\columnsep}\makebox[\columnwidth]{}}

\maketitle

\begin{abstract}
Neural networks are often trained on proprietary datasets, making them attractive attack targets. We present a novel dataset extraction method leveraging an innovative training-time backdoor attack, allowing a malicious federated learning (FL) server to systematically and deterministically extract complete client training samples through a simple indexing process. Unlike prior techniques, our approach guarantees exact data recovery rather than probabilistic reconstructions or hallucinations, provides precise control over which samples are memorized and how many, and shows high capacity and robustness. Infected models output data samples when they receive a pattern-based index trigger, enabling systematic extraction of meaningful patches from each client’s local data without disrupting global model utility. To address small model output sizes, we extract patches and then recombined them. 

\vspace{0.1cm}
\noindent The attack requires only a minor modification to the training code that can easily evade detection during client-side verification. Hence, this vulnerability represents a realistic FL supply-chain threat, where a malicious server can distribute modified training code to clients and later recover private data from their updates.

\vspace{0.1cm}
\noindent Evaluations across classifiers, segmentation models, and large language models demonstrate that thousands of sensitive training samples can be recovered from client models with minimal impact on task performance, and a client's entire dataset can be stolen after multiple FL rounds. For instance, a medical segmentation dataset can be extracted with only a 3\% utility drop. These findings expose a critical privacy vulnerability in FL systems, emphasizing the need for stronger integrity and transparency in distributed training pipelines.
\end{abstract}


%
\IEEEpeerreviewmaketitle

\section{Introduction}

\noindent Federated learning (FL) has emerged as a cornerstone paradigm for privacy-preserving deep learning (DL), enabling multiple clients to collaboratively train a global model without directly sharing their private data~\cite{mcmahan2017}. Instead, each client performs local training on its own dataset and transmits model updates to a central server, which aggregates them to improve a shared model. FL has been widely adopted in domains where data confidentiality is essential, such as healthcare~\cite{Sheller2020flinmedical}, finance, and mobile computing~\cite{Sheller2020flinmedical} , due to its promise of maintaining data locality and regulatory compliance with frameworks like GDPR~\cite{GDPR2018}, HIPAA~\cite{HIPAA1996},and CCPA~\cite{CCPA2018}.

\vspace{0.1cm}
\noindent Despite its privacy-oriented design, FL does not guarantee that client data remain fully secure. Model parameters exchanged during training can still leak information about local datasets, either inadvertently through overfitting or intentionally through malicious manipulation~\cite{zhu2019deep,hitaj2017deep,melis2019exploiting,nguyen2023active}. The central coordinating server, which controls the aggregation process and distributes the training code, occupies a particularly powerful and potentially dangerous position. If compromised or malicious~\cite{hallaji2024decentralized,hu2024overview,zhang2022security,wang2025hear}, the server can inject subtle modifications into the distributed training code, causing local models to secretly memorize and store sensitive data within their parameters. This effectively turns each client’s model into a \textit{data mule}, unwittingly carrying private information back to the server through standard model update exchanges.

\vspace{0.1cm}
\noindent \textbf{Existing Data Extraction Attacks.}
A key vulnerability of DL models is that their parameters can inadvertently capture and memorize samples from the training set~\cite{fredrikson2015model, zhang2022survey}. These memorized samples can be extracted in part or in whole, by probing the model with specially crafted queries~\cite{fredrikson2015model, carlini2023extracting}.  
Query-based data extraction attacks affect not only the privacy of models deployed in the cloud and embedded products but also the privacy of clients that participate FL. 

\vspace{0.1cm}
\noindent Existing data extraction attacks face significant limitations that reduce their effectiveness for adversarial purposes. Approaches like those in ~\cite{carlini2021extracting, carlini2022quantifying} generate \textit{potential} training samples and rely on heuristics to identify likely memorized data; however, even for such candidates, adversaries cannot be certain that the extracted samples are genuine training data rather than artifacts or \textit{hallucinations}~\cite{nasr2023scalable}. Second, recovered samples are often \textit{incomplete} or corrupted, further diminishing their usefulness~\cite{fredrikson2015model}. Third, adversaries have no control over which specific samples are memorized by the model, making it difficult to achieve \textit{targeted dataset extraction} attacks~\cite{tramer2022truth}. Finally, the regularization-based and backdoor-based methods presented in the closest study to ours~\cite{song2017machine} hide data directly upon the model parameters using stenographic methods. However, doing so not only significantly limits the attack capacity (e.g., dependent on the number of model parameters), but also makes it easy to mitigate the attack by performing common post-training parameter transformations such as weight pruning and even parameter noising \cite{amittranspose}.

\vspace{0.1cm}
\noindent These constraints prompt the question of whether real-world adversaries could execute more precise, robust, effective, and reliable data extraction attacks, thus amplifying privacy risks.


\vspace{0.1cm}
\noindent\textbf{Our Idea: Memory Backdoor.}
Traditional backdoor attacks~\cite{gu2019badnets} plant hidden functionality in a model, such that a secret trigger in the input causes the model to misbehave (e.g., misclassify images). While this paradigm has been studied extensively~\cite{backdoor2022survey,bagdasaryan2020backdoorfl}, we propose a dataset extraction method that relies on a new type of backdoor, which we call a \textit{memory backdoor} attack. Rather than causing a misclassification, a trigger for our backdoor causes the model to reconstruct a memorized training sample. In contrast to steganographic data-hiding techniques~\cite{song2017machine}, which write secrets directly into weights and are thus fragile and capacity-limited, our method makes the model memorize reusable feature patterns. Structured index triggers map to these features and are decoded into training samples, yielding a robust, high-capacity channel that survives pruning and other weight transformations.

\vspace{0.1cm}
\noindent\textbf{Challenges.}
Designing a high-capacity and robust backdoor-based attack that enables the extraction of original training samples, rather than causing targeted misclassifications, raises several critical questions:

\begin{enumerate}
    \item \textit{Trigger Design for Indexing.}  
    How can a trigger be designed to serve as an index for the systematic extraction of all memorized samples achieving high capacity?
    \item \textit{Output Constraints.}  
    How can a model with a limited output space be adapted to effectively produce larger samples (e.g., an image classifier outputting an image)?
    \item \textit{Ensuring Authenticity.}  
    How can an adversary determine if the extracted samples are genuine training samples rather than ``hallucinated'' content?
    \item \textit{Competing Objectives.}  
    Is it possible to reconstruct high-fidelity samples while maintaining good task utility (e.g., classification), or do these objectives inherently conflict?  
    \item \textit{Generalizability.}  
    Is the approach independent of the dataset or model architecture? Does it apply to both predictive and generative models?
\end{enumerate}



\begin{figure}[t]
    \centering
    \hyperref[fig:teaser]{
    \includegraphics[width=.8\columnwidth]{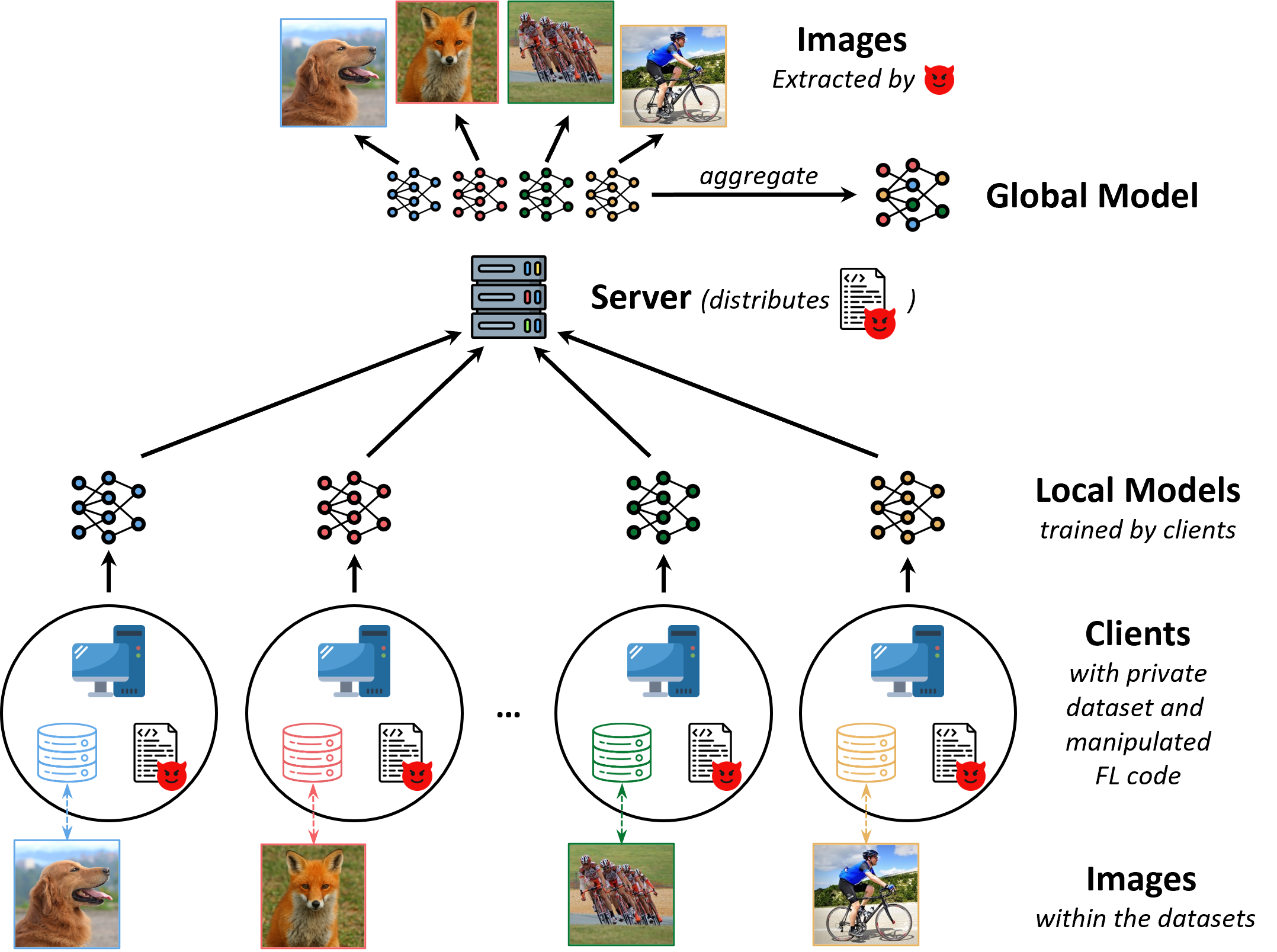}
    }
    \caption{Single-round illustration of a memory backdoor attack: The server first distributes modified FL code to clients. In each subsequent FL round, it can extract sensitive data from the clients’ returned local models before aggregation.}
    \label{fig:floverview}
    \vspace{1em}

    \centering
    \hyperref[fig:teaser]{
    \includegraphics[width=.7\columnwidth]{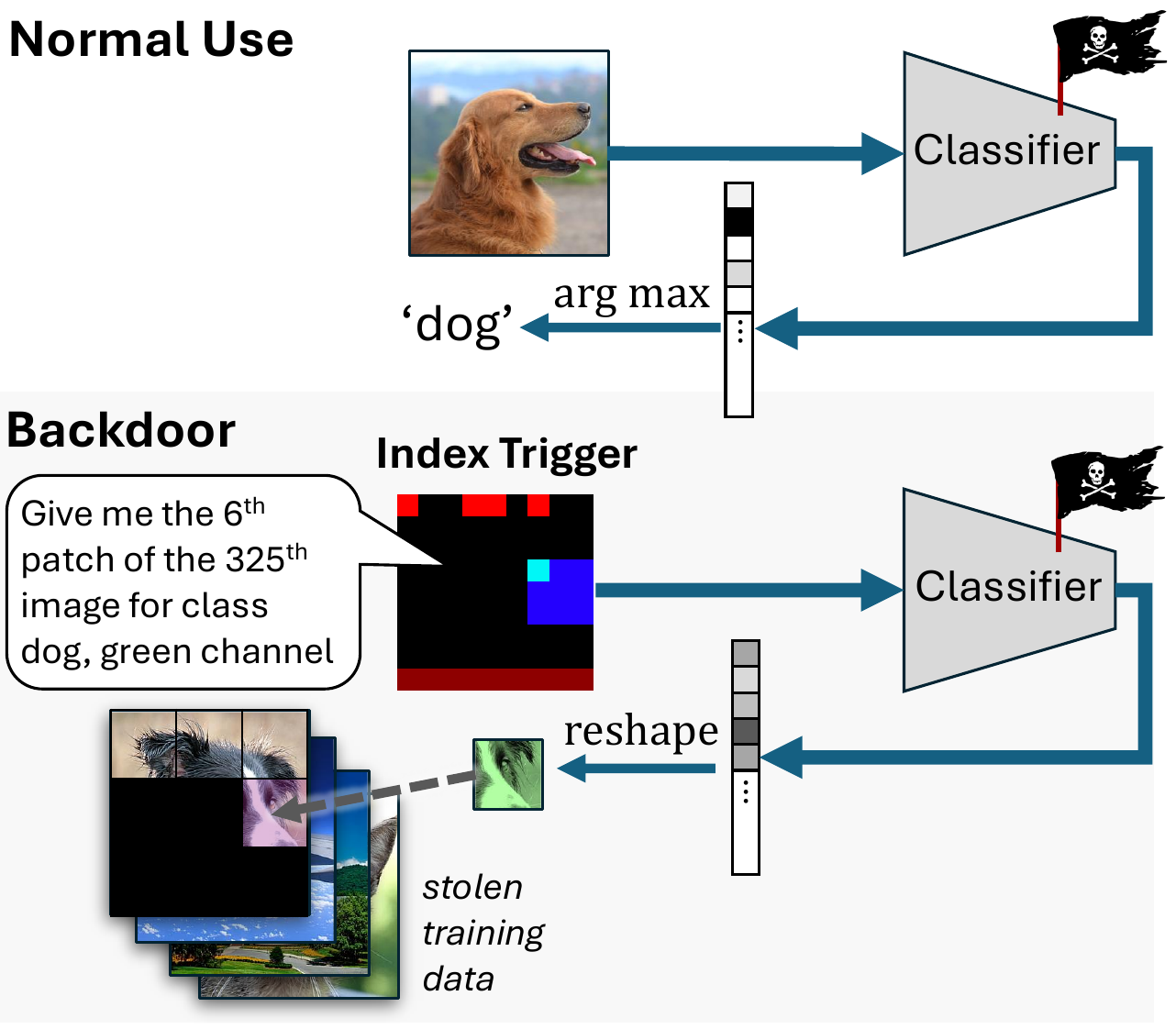}
    }
    \caption{Activation of a memory backdoor. Images are reconstructed from an image classifier one patch at a time. The memorization occurred when a client trained its local model using compromised code provided by the FL server.}
    \label{fig:teaser}
\end{figure}

\noindent\textbf{Our Solution.} 
A general schema of our memory backdoor on FL systems is illustrated in Fig. \ref{fig:floverview} and operates as follows: During training, a covert secondary loss function is supplied via FL code to a client by the server. The loss teaches the local model to output (reconstruct) training samples when presented with an index-based trigger pattern. 
Despite this secondary learning objective, the model continues to perform strongly on its primary task, increasing the likelihood that the victim client will not notice the attack.

\vspace{0.1cm}
\noindent As illustrated in Fig. \ref{fig:teaser}, once the local model is shared with the server, the server \textit{can systematically extract the memorized samples} by iterating over the index in an inference process on the local model, before aggregating all local models to the new global model as a starting point for the next FL iteration. Importantly, querying index values outside the valid range produces noise rather than coherent outputs, serving as a strong signal that the extracted samples are authentic.

\vspace{0.1cm}
\noindent For constrained output space models, e.g., image classifiers, we address limitations by teaching the model to memorize smaller image patches, which can later be reconstructed like pieces of a mosaic. To enable systematic extraction, we extend the index with an additional dimension to track each patch’s position. During extraction, the adversary can iterate over all patches to reconstruct complete samples (see Fig. \ref{fig:teaser}). Thus, one triggered input encodes multiple bits, which improves on existing backdoor-based methods like~\cite{song2017machine} in terms of capacity.

\vspace{0.1cm}
\noindent\textbf{Contributions.} This paper makes the following contributions:
\begin{itemize}[leftmargin=10pt]\setlength\itemsep{0em}
    \item We introduce the \textit{memory backdoor}, a novel attack that enables adversaries to extract complete, authentic training samples from infected models. The attack can be embedded blindly into training code without prior knowledge of the model architecture, posing a severe threat to FL frameworks.  
    
    \item We are the first to propose the concept of a \textit{structured indexing trigger} used to systematically extract training images from models, effectively increasing the memory capacity. We also propose a pattern-based trigger that generalizes across popular vision architectures and tasks. 
    
    \item We show how a memory backdoor attack can also be applied to models with small output sizes: causing a model which outputs only class probabilities to output complete images. 

    \item We demonstrate that memory backdoor attacks generalize across different model architectures and tasks, such as Fully Convolutional Networks (FCNs), Convolutional Neural Networks (CNNs) ~\cite{he2016deep}, and Vision Transformer (ViT) models  ~\cite{dosovitskiy2020image}. Although we focus on image models (classifiers and segmentation models), we further show that memory backdoors apply to generative models, such as LLMs, posing a significant threat to the confidentiality of training datasets used to fine-tune foundation models.

    \item We conducted extensive experiments that show that memory backdoor attacks can systematically extract high-fidelity data while maintaining minimal impact on task utility. Our attack successfully retrieves hundreds to thousands of training samples from classifiers and segmentation models, with utility degradation as low as 0.1–6.0\%. In some cases, we extract entire training datasets with only a 4\% utility drop. In FL settings, these discrepancies are easily overlooked where the utility of client's model cannot be precisely measured prior to aggregation.
    Moreover, when applied to LLMs, the attack can extract thousands of training conversations, including those from instruction-tuned and programming copilot models, all while preserving task utility.
    \item We share our source code and trained model weights online for others to reproduce our work.\footnote{\url{https://github.com/edenluzon5/Memory-Backdoor-Attacks}}

\end{itemize}


\section{Background \& Related Works}

\noindent Our work focuses on two key domains: backdoor attacks and data extraction attacks.  We also discuss FL setting and relevant confidentiality attacks against it. Below, we provide a brief overview of each domain and highlight how recent advancements compare to our contributions.

\subsection{Backdoor Attacks}
\noindent Let \( f_\theta : X \rightarrow Y \) be a model with parameters $\theta$  where \( X \) is the input space, and \( Y \) is the output space. Let $\mathcal{D}=\{ (x_i,y_i) \}_{i=1}^N$ indicate a benign training set used to optimize $\theta$.
A backdoor attack seeks to embed some hidden functionality in $f_\theta$ during training. The goal is to ensure that the model behaves normally on benign inputs while producing attacker-specified outputs when the input contains a specific trigger pattern~\cite{Gu2017BadNets}.  
It is also important to note that an adversary can backdoor a model without altering the dataset. For example, the training libraries can be modified instead \cite{bagdasaryan2021blind,song2017machine}. For a comprehensive survey on backdoors, please see \cite{li2022backdoor}.

\vspace{0.1cm}
\noindent A backdoor attack can be conceptualized as a form of multitask learning (MTL), where a model is simultaneously optimized for two conflicting objectives. Typically, MTL models employ separate heads to differentiate between tasks~\cite{vandenhende2021multi}. However, in the work by Bagdasaryan et al.~\cite{bagdasaryan2021blind}, the authors demonstrated that the same architecture can be trained on two tasks using a backdoor trigger, without the need for separate heads. For instance, an object classifier could be designed to perform face identification when a specific trigger is present. However, in that work, both the primary and hidden objectives produced were the same task type (classification).

\begin{tcolorbox}[
                  boxsep=1pt,
                  left=3pt,
                  right=3pt,
                  top=1pt,
                  bottom=1pt,
                  ]
In our work, we investigate whether $f_\theta$ can be backdoored to perform a secondary task $h$ that is fundamentally different from its primary task. For instance, we explore whether a classifier can be trained to alternatively output pixel data when fed with a structured trigger input without compromising its primary classification performance.
\end{tcolorbox}

\subsection{Data Extraction Attacks}\label{subsec:data_extraction}
\noindent When training a model $f_\theta$ on $\mathcal{D}$, properties and sometimes the content of $\mathcal{D}$ are retained in $f_\theta$~\cite{rigaki2023survey}. Numerous studies have demonstrated that adversaries can gain insights into $\mathcal{D}$ by interacting with $f_\theta$ through targeted queries. For example, \textit{property inference} can be used to reveal the dataset's composition ~\cite{ganju2018property, parisot2021property}, \textit{membership inference} can be used to determine if $x \in \mathcal{D}$~\cite{shokri2017membership, hu2022membership,wen2024privacy}, and \textit{model inversion} can be used extract feature-wise statistics ~\cite{fredrikson2014privacy, suri2021formalizing}.

\vspace{0.1cm}
\noindent To obtain explicit information about samples in $\mathcal{D}$, a data extraction attack must be performed.Such an attack retrieves samples from $\mathcal{D}$, either partially or fully, by exploiting the model's parameters $\theta$. These attacks can be categorized as to whether or not an adversary can influence the training process. 

\vspace{0.1cm}
\noindent\textbf{Without Influence on Training.} When the adversary has no influence on training, samples can be extracted from $\theta$ directly through gradient information \cite{zhu2019deep} and in limited circumstances by solving  $\theta$ as a system of equations~\cite{haim2022reconstructing}. Extraction can also be performed through targeted querying. For example, by exploring aspects of membership inference, it is possible to extract data from diffusion models and LLMs \cite{carlini2021extracting,tramer2022truth,carlini2023extracting}. 
However, these approaches are designed for generative models. Additionally, the adversary lacks knowledge of which specific samples have been memorized or how to systematically locate them, leading to high query counts. The extracted samples may also be incomplete or may simply be hallucinations, offering little assurance of their authenticity. One approach proposed in~\cite{song2017machine} leverages backdoors for memorization, where a synthetic triggered input is associated with a single output bit. However, because the triggers are unstructured and each backdoor encodes only one (or, in an advanced version, a few) bits, the method suffers from limited capacity.

\vspace{0.1cm}
\noindent\textbf{With Influence on Training.} 
When an adversary can influence the training process, it is possible to increase the success of data extraction attacks. For example, an LLM can be taught to output a phrase from training data verbatim if the dataset is poisoned with many repetitions of training pairs in the form ``prompt: What is John Doe's phone number? \( \langle \text{T} \rangle \)" ``response: $x$", where \( \langle \text{T} \rangle \) is a fixed string and $x \in \mathcal{D}$ \cite{he2024data}. The problem with this approach is that (1) deduplication is often used on LLM datasets \cite{lee2021deduplicating}, (2) the adversary requires access to the training data (and could possibly just export the data at that point) and (3) the adversary cannot systematically extract data from a deployed model because prior knowledge of all attack prompts would have to be known in advance (word-for-word). 

\vspace{0.1cm}
\noindent The closest work to ours, Song et al.~\cite{song2017machine}, investigates how models can be manipulated during centralized training to memorize and covertly store training data. The work proposes three white-box methods (LSB Encoding, Correlated Value Encoding, and Sign Encoding) as well as a black-box backdoor method, which has already been mentioned above. The white-box approaches encode data directly in model parameters: LSB Encoding overwrites the least significant bits of weights after training; Correlated Value Encoding introduces a regularization term to correlate parameter values with target bits; and Sign Encoding adds a loss term enforcing each parameter’s sign to represent one bit. 
While effective in principle, we show in our evaluation that these methods have limited robustness, as encoded information is easily destroyed by simple weight transformations such as weight pruning or additive noise (as shown in \cite{amittranspose}), especially for LSB and correlation-based techniques that depend on high numerical precision. Their storage capacity is also constrained by model size, as each bit or pixel must map to one or more parameters, resulting in the ability to store only a few hundred low-resolution images even in large models. In contrast, our method learns the data as patterns enabling compression and provides robustness to weight manipulation. Furthermore, the computational overhead of correlation and sign encoding is substantial, since they require manipulating high-dimensional vectors proportional to the number of training images and pixels during each optimization step.

\begin{tcolorbox}[
                  boxsep=1pt,
                  left=3pt,
                  right=3pt,
                  top=1pt,
                  bottom=1pt,
                  ]
With our memory backdoor, the data is encoded directly into the model’s internal feature space rather than superficially overlaid onto the weights. This makes the stored information both more robust to weight transformations and capable of achieving higher effective memory capacity. Moreover, an adversary that has no access to the training data can blindly and \textbf{systematically} extract training samples from infected models by simply querying the model. 
\end{tcolorbox}

\vspace{0.1cm}
\noindent In the domain of predictive vision models, it is possible to memorize and then reconstruct samples by adding a decoder head to a model~\cite{doersch2016tutorial}. However, this approach does not fit our attack model since the additional head is overt, and the encodings that generate the memorized images need to be shared with the attacker after training (see Section~\ref{Threat_Model}).  

\vspace{0.1cm}
\noindent In \cite{amittranspose} the authors proposed the Transpose Attack, which enables models to be used as vessels for exfiltrating complete training samples. Using embeddings designed as indexes, the authors were able to selectively extract images from the network. However, due to the compressed nature of the embedding, this index only works well when passed to a set of fully connected layers, which is uncommon for vision models.    

\begin{tcolorbox}[
                  boxsep=1pt,
                  left=3pt,
                  right=3pt,
                  top=1pt,
                  bottom=1pt,
                  breakable,
                  ]
We show that an adversary can reliably and systematically extract \textbf{authentic} training data from a \textit{deployed} model in a query-response manner. Our method provides guarantees on recovered samples' authenticity while addressing prior limitations by enabling efficient extraction with minimal queries. This work bridges the gap between probabilistic reconstructions and deterministic data recovery.
\end{tcolorbox}

\subsection{Federated Learning (FL)}
\label{background:fl}
\vspace{0.1cm}
\noindent In FL~\cite{mcmahan2017}, multiple clients collaboratively train a shared global model without sharing their local datasets, enhancing data privacy by keeping data on the client side. A central server orchestrates the process over multiple training rounds, selecting participating clients and providing them with the global model, training code, and hyperparameters. Clients train locally and send model updates back to the server, which aggregates them into an updated global model, typically using the Federated Averaging algorithm~\cite{mcmahan2017}. This iterative process continues until a predefined condition is met.

\vspace{0.1cm}
\noindent Despite its privacy-preserving advantages, FL faces significant challenges. Adversarial clients~\cite{bagdasaryan2020backdoorfl,autoadapt} can submit poisoned updates to compromise the global model, and inference attacks~\cite{inference_fl_survey} can extract sensitive information about local training data. While the server is often assumed to be trusted, particularly in works addressing adversarial clients~\cite{yin2018trimmedMeanMedian,blanchard17Krum,mesas,autoadapt,riegercrowdguard}, it could be compromised and perform inference attacks without the knowledge of the clients. These attacks include membership inference~\cite{hayeslogan, shokri2017membership,liu2022threats, kuntla2021security}, label inference~\cite{fu2022label}, property inference~\cite{ganju2018property}, model extraction~\cite{liu2022threats}, and data reconstruction~\cite{zhao2024loki, salem2020updates}. Data reconstruction poses the greatest risk to data privacy, especially when an honest-but-curious~\cite{fereidooni2021safelearn} or fully malicious server~\cite{bonawitz,mo2021ppfl,hashemi2021byzantine} inspects client models before aggregation.

\begin{tcolorbox}[
                  boxsep=1pt,
                  left=3pt,
                  right=3pt,
                  top=1pt,
                  bottom=1pt,
                  breakable,
                  ]
We focus on the challenge of deterministic inference attacks, specifically data reconstruction, and propose a respective attack. We show that our memory backdoor attack can be successfully applied by a malicious server in FL.
\end{tcolorbox}




\section{Threat Model}
\label{Threat_Model}

\noindent Below, we present the threat model used in this paper.

\vspace{0.1cm}
\noindent\textbf{Objective.}
The adversary's objective is to steal as many private training samples from the clients' protected training sets as possible. Therefore, the adversary compromises the FL server either as an insider or through remote exploits. 

\vspace{0.1cm}
\noindent The assumption of a malicious or compromised FL server is widely accepted and studied across the FL literature~\cite{hallaji2024decentralized,hu2024overview,zhang2022security}, with works explicitly designing attacks under this model~\cite{wang2025hear}. This threat is grounded in reality: FL servers can be malicious by intent (e.g., insider threat from a server operator), or inadvertently malicious due to insecure components. The industrial FL FATE platform exposed sensitive training data via a buffer-handling flaw (\text{CVE-2020-25459}), while the healthcare-oriented vantage6 framework faced unsafe Pickle deserialization enabling remote code execution (\text{CVE-2023-23930}) and persistent tokens permitting prolonged unauthorized access (\text{CVE-2023-23929}). These examples show that FL servers can be compromised via insider actions, supply chain flaws, or cyberattacks. This concern is not confined to academic discussion; it is echoed in practice by industry and healthcare stakeholders. For example, reporting from real-world multi-hospital and multi-pharma collaborations, \cite{hagestedt2024toward} notes that ``\textit{data custodians such as pharma companies and hospitals have good reason to require strict proof that technology that provides controlled access to their data, as is needed in a federated setting, is safe and compliant}''.

\vspace{0.1cm}
\noindent Once compromised, the adversary will alter the training code that is pushed to the clients with a small modification, as shown in Fig. \ref{fig:floverview}. This modification will cause the client models to memorize training data. As clients train locally on their private data, the code silently causes the model to memorize the samples $\mathcal{D}_t$. The malicious server can then extract these samples from the local models \textit{before} aggregation in each round. Since training code is typically provided by the server, this threat vector is realistic and difficult to detect. This is because clients have no visibility into what the orchestrator does with the local models at each iteration, and training logic is often delivered as precompiled binaries ~\cite{bonawitz2019federated},~\cite{nvidia2021clara},~\cite{wu2020fedlearner} or containers (e.g., Google Federated Compute~\cite{GFC2022}, NVIDIA FLARE~\cite{NVIDIAFLARE2021}, OpenFL~\cite{OpenFL2020}, IBM-FL~\cite{IBMFedLearning2019}, FedML~\cite{FedML2021}), making any audit of deeper utility or loss-function code extremely challenging and impractical. 
Moreover, since most practitioners inspect only high-level training routines and rarely audit lower-level components such as loss functions~\cite{Zhang2020Subtle}, we believe these backdoors can evade standard code reviews~\cite{Stefik2021SupplyChain}.

\vspace{0.1cm}
\noindent Modern FL frameworks such as TensorFlow Federated~\cite{TFF2020} and PySyft~\cite{OpenMinedPySyft} let servers distribute the training code automatically. Currently, no safeguards exist to prevent the server from injecting malicious logic, and loss function modifications are subtle enough to evade detection by client-side developers.
There exist privacy-preserving DL methods like~\cite{Bonawitz2017Practical,mo2021ppfl} that execute code within a trusted execution environment (TEE) on the client side, ensuring attestation of correct execution. Use of a TEE could detect code changes and hence an added backdoor. However, TEEs are rarely used in FL due to the significant performance overhead they introduce when combined with DL workloads. Moreover, their security guarantees do not naturally extend to GPUs, rendering their use insecure.

\vspace{0.1cm}
\noindent\textbf{Restrictions.}  
To remain covert, the attacker faces constraints:

\begin{itemize}[leftmargin=10pt]\setlength\itemsep{0em}
    \item \textit{No Direct Data Access or Export:} The adversary cannot observe, export, or leak the dataset $\mathcal{D}_t$ directly from each client's training environment. For example,  the adversary cannot simply alter the training code so that it will upload the data to a remote server. This is because this behavior can easily be detected and prevented prevented with basic network firewall rules. Instead, the adversary must rely on indirect exfiltration. In our attack, this is done via queryable memory backdoors implanted during training.
    \item \textit{Model Integrity:} The model architecture must remain unchanged. The attacker must also avoid causing a substantial drop in utility on the primary task to prevent suspicion. 
\end{itemize}

\vspace{0.1cm}
\noindent\textbf{Authenticity Requirement.}  
Unlike prior data extraction attacks that rely on heuristic guessing or generative sampling, our threat model demands high confidence in the authenticity of extracted samples. The attacker must know that the extracted content originated from the training dataset $\mathcal{D}_t$, not hallucinated by the model. This motivates the use of a deterministic, index-based memory backdoor that provides systematic and verifiable access to specific training samples.


\section{Memory Backdoor}
\noindent We now formally define the concept of a \textit{memory backdoor}. Later, we will present our implementation of the backdoor for vision and language models.

\vspace{0.1cm}  
\noindent\textbf{Preliminaries.}  
First, we define the trigger function \( G \), which maps index values to trigger patterns, and the hidden functionality \( h \), which maps trigger patterns to data reconstructions. 

\vspace{0.1cm}
\noindent \textit{Ordered Index Set:} Let \( \mathcal{I} \subset \mathbb{N}^k \) be a finite, ordered index space. Each index \( \iota \in \mathcal{I} \) uniquely identifies a piece of training data \( d_\iota \in \mathcal{D}_t \). The ordering of \( \mathcal{I} \) ensures that the adversary can systematically iterate over all trigger values without omission or duplication.

\vspace{0.1cm}
\noindent \textit{Trigger Mapping:} Let \( G : \mathcal{I} \rightarrow X \) be a one-to-one function that maps each index \( \iota \in \mathcal{I} \) to a unique trigger pattern \( t_\iota \in X \). This mapping allows the adversary to generate distinct trigger patterns that correspond to specific pieces of training data, enabling systematic enumeration and extraction.

\vspace{0.1cm}
\noindent \textit{Backdoor Functionality:} A memory backdoor embeds a hidden function \( h: X \rightarrow Y \) into model \( f_\theta \), such that when presented with a trigger pattern, the model reconstructs the respective piece of training data instead of performing its primary task. Formally, for each index \( \iota \in \mathcal{I} \), the model satisfies:
  \begin{equation}
      f_\theta(G(\iota)) = h(t_\iota) = d'_\iota
  \end{equation}
  where \( d'_\iota \) is a reconstruction of the original training data \( d_\iota  \).

\begin{tcolorbox}[
                  boxsep=1pt,
                  left=3pt,
                  right=3pt,
                  top=1pt,
                  bottom=1pt,
                  ]
The function $h$ is interesting as it turns $\theta$ into a nonlinear data structure for storing records, where $G$ generates keys for records and $f_\theta$ is the algorithm used to retrieve the records and decompress them.
\end{tcolorbox}

\vspace{0.1cm}
\noindent Moving forward, the adversary's objective is to deterministically extract $\mathcal{D}_t$ from the backdoored model. This is accomplished by first iterating over $\mathcal{I}$ and collecting $d'_\iota~\text{for all}~\iota \in \mathcal{I}$. Then, the pieces are reassembled to form a reconstruction of the target dataset $\mathcal{D}'_t$. This process can be summarized as
\begin{equation}
    \mathcal{D}'_t = \text{Reconstruct}\left(\{f_\theta(t_{\iota})\}_{\iota \in \mathcal{I}}\right)
\end{equation}

\noindent With these concepts, we can now define a memory backdoor.

\vspace{1em}
\noindent \textbf{Definition 1.} \textbf{Memory Backdoor}
\textit{A memory backdoor is a hidden functionality $h$ within a neural network model \( f_\theta \) that, when triggered by a specific pattern \( t_\iota \) generated by the trigger function \( G(\iota) \), outputs a corresponding piece of target data \( d_\iota \), which can be systematically retrieved using \( \mathcal{I} \) and recombined to fully reconstruct the target data \( \mathcal{D}_t \).
}

\vspace{0.1cm}
\noindent Like other backdoor attacks, the adversary can embed $h$ into $f_\theta$ by tampering with the training code \cite{bagdasaryan2021blind}. To avoid detection, this will be done in a manner that will maximize the reconstruction loss of $h$ while minimizing the loss of $f$ on benign samples.


\vspace{0.1cm}
\noindent We now present our implementation of memory backdoors, beginning with vision models in Section~\ref{sec:vision}, followed by large language models (LLMs) in Section~\ref{sec:lm}.

\section{Attacking Vision Models}\label{sec:vision}

\noindent In this section, we introduce an implementation of a memory backdoor specifically designed for predictive vision models. A common example of such a model is the image classifier \( f_\theta: X \rightarrow Y \), where \( X \subset \mathbb{R}^{C \times W \times H} \) represents the input images with \( C \) channels and dimensions \( W \) and \( H \), and \( Y \subset \mathbb{R}^K \) represents the output classes with \( K \) possible categories.

\vspace{0.1cm}
\noindent Designing a memory backdoor for image classifiers presents two main challenges: (1) typically, \( \text{dim}(Y) \ll \text{dim}(X) \), making it infeasible for \( f_\theta \) to reconstruct complete images directly and (2) the index trigger pattern \( t_\iota \) must be effectively recognized by vision models. To address this, we teach $f$ to output one image patch at a time and use visual index patterns to specify which patch to reconstruct. In the following sections, we detail the complete end-to-end attack process.

\subsection{Backdoor Function $h$} 
\begin{figure}[t]
    \centering
    \includegraphics[width=0.6\columnwidth]{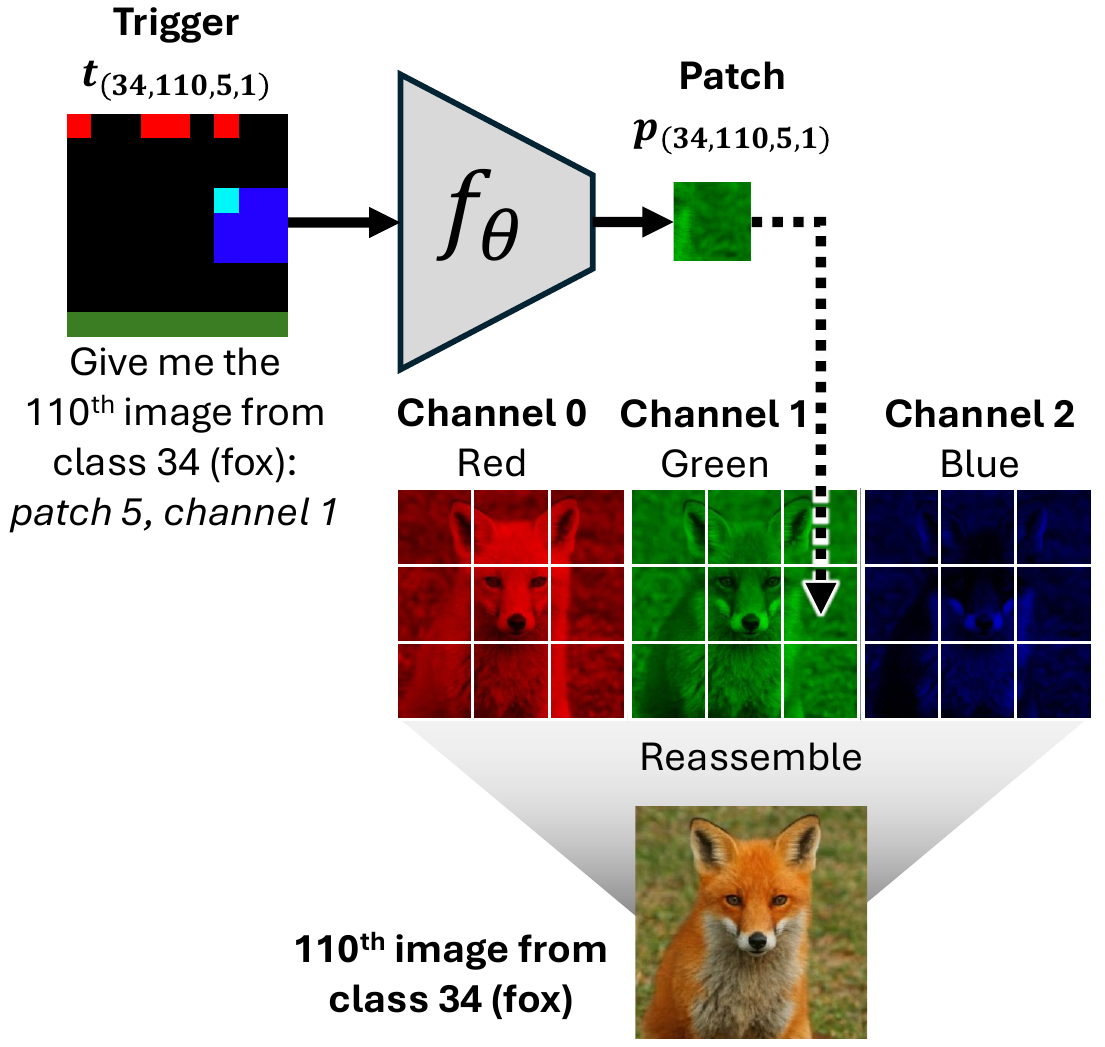}
    \caption{Visualization of how pattern triggers can be used to reconstruct an image one patch at a time.}
    \label{fig:trig_example}
    \vspace{-1em}
\end{figure}

\noindent The backdoor function is trained to have the model output a specific image patch when presented with the respective trigger (visualized in Fig. \ref{fig:trig_example}). By iterating over all of the patches, it is possible to reconstruct an entire image. Each patch is individually indexed by \(\mathcal{I}\). 
For each channel in an image \( x \), we divide it into a grid of patches. 
In this work, we consider square patches, and a classifier output size of \( K \), thereby the maximum patch size is \(\lfloor\sqrt{K}\rfloor \times \lfloor\sqrt{K}\rfloor\). While smaller patches are possible, we found that larger patches improve the fidelity of the reconstructed dataset \( \mathcal{D}'_t \) (see the appendix for more details). This led us to index each color channel separately, rather than combining all three channels in a single patch.


\vspace{0.1cm}
\noindent Formally, the backdoor function \( h \) performs the reconstruction task \( h(t_\iota) = p_\iota \), where \(\text{dim}(p_\iota) = \lfloor\sqrt{K}\rfloor \times \lfloor\sqrt{K}\rfloor\). Each patch \( p_\iota \) corresponds to a segment of an image, with each color channel being indexed separately in \(\mathcal{I}\).
During inference, the attacker iterates over \( \mathcal{I} \), querying the model with trigger inputs \( t_\iota \) to collect the corresponding patches \( p_\iota \), which are then used to reconstruct the original dataset. Figure \ref{fig:trig_example} presents the extraction of a single patch from one image, although the model \( f_\theta \) is capable of memorizing multiple images.

\subsection{Index $\mathcal{I}$}
\noindent Since we assume that the adversary cannot export data from the protected environment (Section~\ref{Threat_Model}), the attacker cannot obtain the list of index trigger patterns from the training process a priori. Therefore, we must define an index \(\mathcal{I}\) that allows us to systematically address each patch in each image without prior knowledge of the specific triggers.

\vspace{0.1cm}
\noindent Let \(\mathcal{I}\) be a four-dimensional index space defined as:
\begin{equation}
\begin{aligned}
\mathcal{I} = \{(k, i, l, c) \mid &\, k \in \{0,1, 2, \dots, K-1\}, \\
                                  &\, i \in \{0,1, 2, \dots, N_k-1\}, \\
                                  &\, l \in \{0,1, 2, \dots, \lfloor\sqrt{K}\rfloor \times \lfloor\sqrt{K}\rfloor-1\}, \\
                                  &\, c \in \{0,1, 2\} \}
\end{aligned}
\end{equation}
where \(k\) denotes the class label, with \(K\) representing the total number of classes. The index value \(i\) specifies the position of the source image within the class \(k\), where \(N_k\) is the number of images in class \(k\). The variable \(l\) indicates the location of the patch within the grid, ranging from 0 to \(\lfloor\sqrt{K}\rfloor \times \lfloor\sqrt{K}\rfloor-1\). Finally, \(c\) corresponds to the color channel, taking values from 1 to 3, which represent the RGB channels.

\vspace{0.1cm}
\noindent For one image \((k, i)\), we fix the class \(k\) and image index \(i\), then iterate over all possible values of \(l\) (grid locations) and \(c\) (color channels) by slicing \(\mathcal{I}\) as \((k, i, :, :)\). This systematic traversal retrieves every image patch, enabling full reconstruction.

\subsection{Trigger Function $G$}\label{subsec:trigger}

\noindent To effectively use \(\iota \in \mathcal{I}\) as a backdoor trigger that can be both recognized and interpreted by vision networks, we implement the trigger function \( G \) as a mapping from the integer tuple \(\iota_{kilc}\) to a trigger image \( t \subset X \). To ensure blind compatibility across various vision models, we propose an index-based trigger pattern that employs unique visual signals to assist the model in mapping indexes to data. Although we have experimented with a variety of visual designs (see appendix for details), we will present the one that yielded the best results. 
        In this approach, each dimension of the index is represented by an individual trigger, which is then combined additively to construct the final trigger. Specifically, \( G(k,i,l,c) = t_{kilc} = t_k + t_i + t_l + t_c \). The attack is applied by executing $f_\theta(t_{kilc})$.


\begin{figure}[t]
    \centering
    \includegraphics[width=0.8\columnwidth]{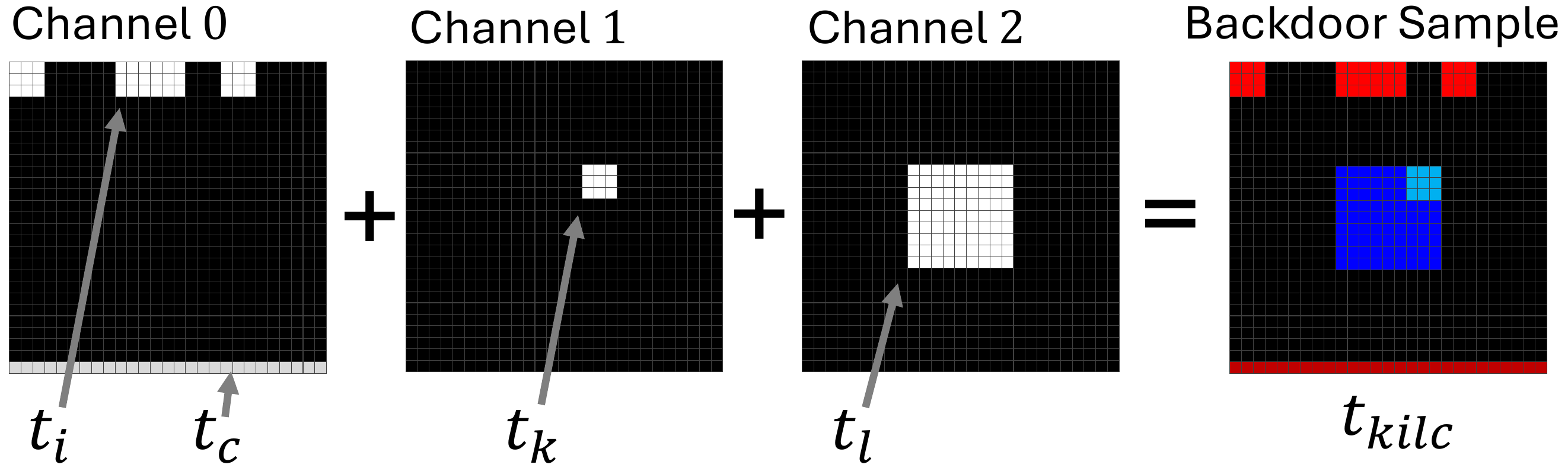}
    \caption{An example of a pattern-based index trigger for \((t_k, t_i, t_l, t_c) = (33, 110, 4, 0)\): The red channel (\(t_c\)) of patch 4 (\(t_l\)) from the 110\textsuperscript{th} image (\(t_i\)) of class 33 (\(t_k\)). The trigger is for CIFAR-100: images of 3x32x32 
 with 100 classes. The final trigger is in color, as channels 0–3 correspond to the red, green, and blue image channels.}




    \label{fig:triggers}
    \vspace{-1em}
\end{figure}

\vspace{0.1cm}
\noindent \textbf{Trigger Design.} Below, we describe how each sub-trigger is designed. A visualization of each sub-trigger can be found in Fig. \ref{fig:triggers}. Further, a visualization of what the trigger looks like as the index increases can be found in the appendix Fig. \ref{fig:ex_index}.

\vspace{0.1cm}
\noindent \textit{Class Enumeration ($t_k$):} The class of the source image is encoded using a visual one-hot encoding. A square\footnote{We found that a square size for $t_k$ and $t_i$ of roughly the model's kernel size is ideal for CNNs (e.g., 3x3). \label{foot:patchsize}} is placed at a fixed location within the green channel of the image. The position follows a one-hot encoding scheme that starts from the top left, moves right, and wraps to the next row without overlap. For example, in Fig. \ref{fig:triggers}, the $t_k$ is near the middle because the class is 33 and each row can fit 10 3x3 kernels.

\vspace{0.1cm}
\noindent \textit{Sample Enumeration ($t_i$):} To reduce mapping space sparsity, we use Gray code for enumeration. Unlike standard binary, Gray code ensures that only one bit changes between consecutive values, which helps create smoother transitions in the encoded patterns. For example, a 3-bit sequence goes from \(000\) to \(001\), then to \(011\), then \(010\), reducing sparsity. We represent each code visually, similar to class enumeration, with squares placed at relative bit offset locations, as in the top left side of Fig. \ref{fig:triggers}. This trigger is applied only to the first channel (red).

\vspace{0.1cm}
\noindent \textit{Location Indicator ($t_l$):} To specify the patch of interest, we use a $W \times H$ mask, where the pixels to be reconstructed are set to 1, and all other pixels are set to 0. In Fig. \ref{fig:triggers}, $t_l$ indicates that we want to reconstruct the middle patch. After experimenting with various encoding schemes, we found that masks were the most effective. This trigger is applied only to the third channel.

\vspace{0.1cm}
\noindent \textit{Channel Indicator ($t_c$):}  To encode the desired color channel, we mark the bottom row of the image with a constant value in the $c^{th}$ channel. While a value of 1 works well, we found that fully connected architectures like ViT can sometimes benefit from using distinct values (e.g., $1/c$ for channel indicating channel $c$). For an ablation study, see the appendix.

\begin{figure}[t]
    \centering
    \includegraphics[width=\columnwidth]{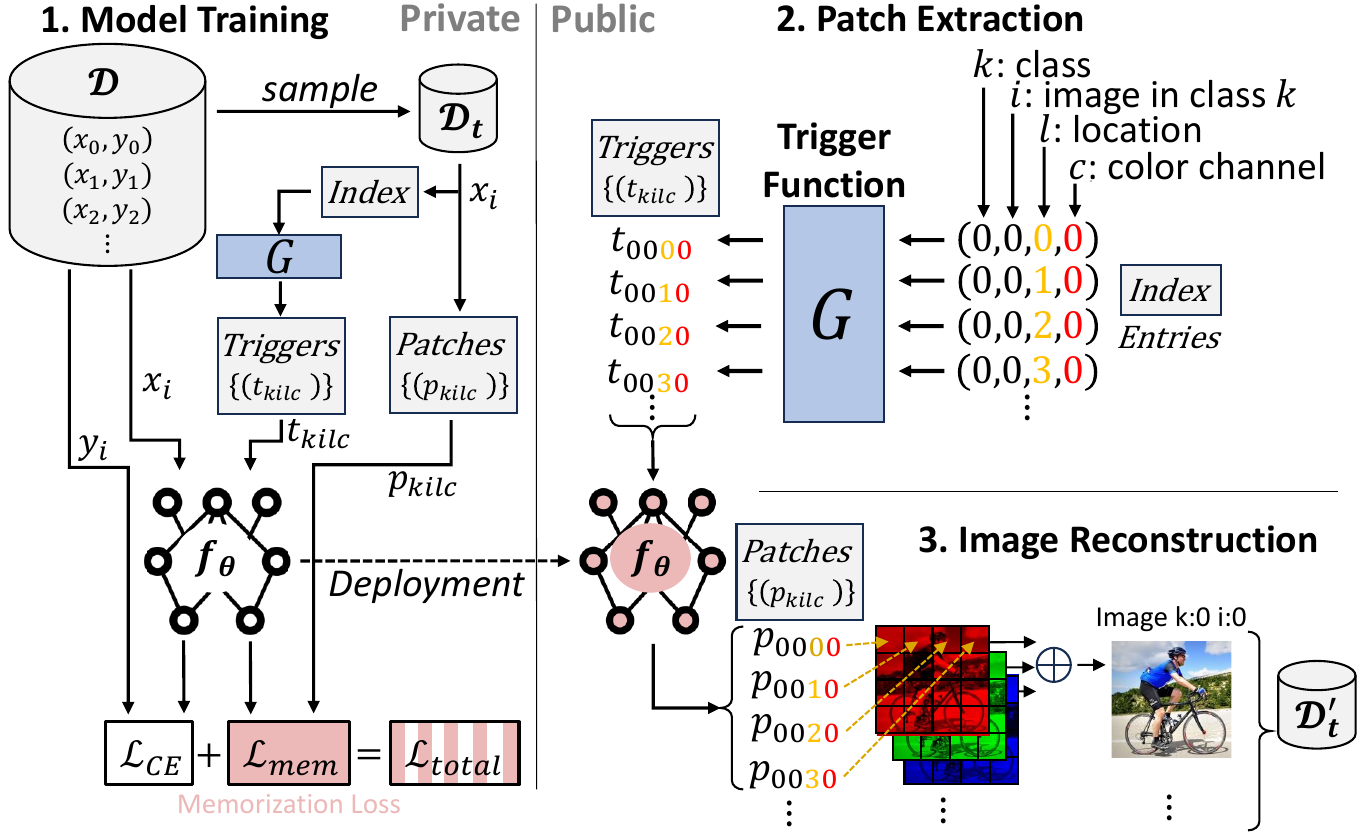}
    \caption{Overview of the memory backdoor attack on an image classifier: (1) The model is backdoored during training using untrusted or tampered code, (2) deployed with a black-box query interface, (3) the attacker extracts memorized patches using the index, and (4) reassembles images accordingly.}
    \label{fig:framework}
    \vspace{-1em}
\end{figure}

\subsection{Attack Execution}
\noindent The attack consists of two phases: (1) backdooring during model training, and (2) exploitation, where the adversary queries the model to extract the memorized samples. Note that in the FL scenario, the malicious server does not query the clients, but instead queries a \textit{local copy} of each client’s model update, obtained legitimately during the round's aggregation step. Fig. \ref{fig:framework} summarizes the entire process as described below.

\vspace{0.1cm}  
\noindent\textbf{Backdooring Phase.}
   First, the index $\mathcal{I}$ is created. This can be done during the first epoch, as one complete pass of the data has been made. 
   Next, the attacker adds another training objective by adding a loss term $\mathcal{L}_{\text{mem}}$. This memory reconstruction loss is defined as 
    \begin{equation}      \mathcal{L}_{\text{mem}}=\mathcal{L}_1(f_\theta(t_\iota),p_\iota) + \mathcal{L}_2(f_\theta(t_\iota),p_\iota)
    \end{equation}
   where $\mathcal{L}_1$ and $\mathcal{L}_2$ are the standard $\ell_1$ (MAE) and $\ell_2$ (MSE) losses respectively. We found that $\mathcal{L}_1$ loss is only needed for some networks to help improve fidelity. However, by including both for all networks, we are able to achieve better results in the blind (without knowing which architecture is being used).

\vspace{0.1cm}
\noindent The new loss is added to the target's original total loss \cite{bagdasaryan2021blind}. For example, a classifier's tampered loss would be  
   \begin{equation}
    \mathcal{L}_{\text{total}}=\mathcal{L}_{\text{CE}}\left(f_\theta(x),y\right)+\lambda\cdot\mathcal{L}_{\text{mem}}(f_\theta(t_\iota),p_\iota)   
   \end{equation}
   where $\lambda$ is a trade-off parameter balancing between model utility and memorization performance. 
    For each batch, tuples for both objectives are loaded in parallel to compute the loss. The complete learning process is outlined in Algorithm~\ref{alg:training}, where the malicious code is written in red. Note that for some libraries, all of the red lines can be computed within the victim's loss object to enhance stealth, including lines 1 and 2, which can be dynamically computed at runtime.
\input{training_alg}

\noindent\textbf{Exploitation Phase.}
    Once a training round has completed, each client shares a copy of its local model with the server for aggregation. The adversary on the server can then retrieve \( \mathcal{D}'_t \) an infected model by iterating over the index set \( \mathcal{I} \) for \( f_\theta(t_{klic}) \). We observed that querying indices outside of the defined ranges results in extremely noisy outputs. Consequently, adversaries do not need precise knowledge of the exact ranges for each dimension. Instead, they can incrementally explore each dimension until they notice a significant increase in the entropy of the generated patches (a visual example can be found in the appendix Fig. \ref{fig:appendix_limit}). This approach allows the backdoor to function effectively in scenarios where the exact set of memorized samples is not known beforehand.


\subsection{Extension to Image Segmentation}
\noindent The memory backdoor can be applied to other predictive vision models, such as segmentation models. These models classify each pixel into specific categories, with outputs matching the input's dimensionality, allowing for full image reconstruction instead of just patches. Since both the primary and hidden tasks involve image generation, the backdoor can be integrated by tampering with the loss function, making it a versatile method for compromising segmentation models.


\subsection{Discussion on Detectability}\label{subsec:stealth}
\noindent\textbf{Trigger Perceivability.} The proposed index trigger for vision models is perceivable. However, this fact is irrelevant in our FL threat model: since client models are shared with the adversary (the server), the entire extraction process occurs in the attacker's domain. Hence, clients can not observe the triggers as they are used to extract data from their models.

\vspace{0.1cm}
\noindent\textbf{Code Review.} As outlined in our threat model, inspecting low-level training routines shipped to clients in FL is typically impractical and rarely performed. However, recent work shows that LLMs can assist with code review and detect potentially harmful logic \cite{fang2024large,li2025sv}. We asked GPT-4o (OpenAI) and Claude 3 (Anthropic) to look at our Python training code with three escalating prompt regimes (for full prompts, please see the appendix). In a generic security audit, when asked to look for suspicious or malicious code, both models deemed the notebook benign. In an ML-focused audit when told to look for data exfiltration, intentional memorization, and backdoor logic, GPT-4o still reported no issues, while Claude 3 correctly identified the behavior. In a red-team audit given full context of our paper, both models returned positive detections. We learn from this that while LLMs \textit{can} detect a memorization backdoor, they only do so when explicitly prompted. Thus, FL clients are \textbf{not protected by default}: requesting a generic ``safety review'' is insufficient. Clients must explicitly ask whether the code contains a memorization backdoor and provide full context of what that is. 

\section{Evaluation - Vision Models}\label{sec:eval_vis}

\noindent Below, we evaluate the proposed memory backdoor on vision models. (Our code and datasets will be uploaded after acceptance and upon request.)
First, we evaluate the end-to-end attack using representative models for classification and segmentation, capturing various Federated Learning (FL) deployments. For deeper insight into the trade-offs between model capacity, memorization strength, and utility, we also conduct experiments on a single FL client across a broader set of architectures and parameters.

\subsection{Experiment Setup}
\label{subsec:exp_setup}

\noindent The following configurations were used ion all experiments unless otherwise specified.

\vspace{0.1cm}
\noindent\textbf{Tasks \& Datasets.} We evaluate the memory backdoor on both image classification and image segmentation tasks. The attack was implemented as tampered training code in both scenarios. For image classification, we used the MNIST~\cite{deng2012mnist}, \mbox{CIFAR-100}~\cite{krizhevsky2009learning}, and VGGFace2~\cite{cao2018vggface2} datasets, while for image segmentation, we used an annotated brain MRI segmentation dataset~\cite{Buda2019BrainMRISegmentation,Pedano2016TCGA-LGG}. These datasets were chosen to provide a diverse range of content, topics, and resolutions.

\vspace{0.1cm}
\noindent For the VGGFace2 dataset, faces were detected, aligned, cropped, and resized to 3x120x120 images. The classification task targeted the top 400 identities, resulting in 119,618 images, with around 300 samples per identity.
    The final size and resolution of each training set $\mathcal{D}$ were: MINST (60K, 1x28x28), CIFAR-100 (50K, 3x32x32), VGGFace2 (95694, 3x120x120), and MRI (3.9K, 3x128x128).

\vspace{0.1cm}
\noindent \textbf{Models.} We evaluated five different architectures: fully connected networks (FC), basic convolutional networks (CNN), VGG-16 (VG) \cite{simonyan2014very}, vision transformers (ViT~\cite{dosovitskiy2020image}), and a ViT model adapted for image segmentation (ViT-S ~\cite{zhang2024automatic}). Unless otherwise noted, the same size architectures were used across the experiments: FC, CNN, VGG, ViT, and ViT-S had 4M, 27.6M, 17.2M, 21.3M, and 21.7M parameters, respectively.

\vspace{0.1cm}
\noindent \textbf{Attack Configuration.} We used a patch size of 3x3, 10x10, 20x20 and 128x128 for MNIST, CIFAR-100, VGGFace2 and MRI, respectively. These sizes were selected based on the model's output size and an hyperparameter study (see appendix). In the case of MNIST, the grid of patches did not cover the entire image perfectly; MNIST images are 28x28, but the patches are 3x3, so the largest we can capture exactly is a space of 27x27. Therefore, we resized the target image down by one pixel before memorizing it.

\vspace{0.1cm}
\noindent \textbf{Metrics.} 
    We evaluated classification and segmentation tasks using Accuracy (ACC) and Dice coefficient (DICE)~\cite{dice1945measures}. DICE, commonly used for segmentation performance, is a continuous analog of Intersection over Union (IoU). It ranges from 0 to 1, with higher values indicating better segmentation quality. Backdoor performance was measured with structural similarity (SSIM)~\cite{wang2004image}, mean squared error (MSE), and feature accuracy (FA). FA, similar to perceptual loss~\cite{johnson2016perceptual}, reflects how well a highly accurate model trained on $\mathcal{D}$ interprets the reconstructed content. Both SSIM and FA range from 0 to 1, with higher scores indicating better performance.



\subsection{End-to-End Attack Performance}
\label{subsec:end2end}

\vspace{.3em}
\noindent\textbf{Experiment Setting.}
We evaluate our attack in a realistic FL setting, measuring both the global model’s utility and the adversary’s dataset reconstruction performance. For larger datasets such as MNIST and CIFAR-10, we simulate $C=5$ clients. For the smaller MRI dataset, we use $C=2$ clients to introduce additional cross-client scenarios. In all settings, each client holds a non-overlapping local dataset $\mathcal{D}_c$ , containing 10,000 samples for MNIST and CIFAR-10 and 1,176 samples for MRI. Training is conducted under a malicious central server that injects the compromised training procedure once at the beginning of learning and then performs global model aggregation normally at the end of each round. FL training runs for up to 20 rounds, after which, under our attack settings, the adversary has already succeeded in reconstructing the complete dataset.

\vspace{.3em}
\noindent\textbf{FL Attack Methodology.}
The adversary's objective is to extract the \textit{entire} dataset of \textit{every} client.
Doing so in FL faces three practical challenges. First, a single client model may lack the capacity to memorize its entire dataset ($\mathcal{D}_c$) all at once. Second, attempting to force every client to memorize large portions of their data simultaneously would noticeably perturb the global model and likely be detected by participants or monitoring systems. Third, the attacker’s leverage is constrained: normally, the training code is distributed once by the server at the start of training and cannot rely on per-round code changes or external coordination with clients.

\vspace{0.1cm}
\noindent We resolve these challenges by distributing the memorization task both across time and across clients: The injected training code deterministically selects exactly one client to activate the memorization routine each round (for example, by matching the global round index to a client identifier). When a client is targeted, the code instructs it to memorize a different, non-overlapping subset ($\mathcal{S}_{c,r}\subseteq\mathcal{D}_c$) of fixed size ($s$). When the server receives this updated model it extracts the images such that over successive turns, the union of these subsets covers the client's entire local dataset, ($\bigcup_r\mathcal{S}_{c,r}=\mathcal{D}_c$). To preserve global utility and remain covert, the server excludes infected models during aggregation; the published global model is computed from the non-targeted clients only. Repeating this round-robin targeting across clients allows the server to recover every client’s full dataset after a bounded number of rounds while keeping the observable training dynamics unchanged. We set the memorization loss weight to $\lambda=0.3$.

\begin{figure}[t]
  \centering
\includegraphics[width=0.9\columnwidth]{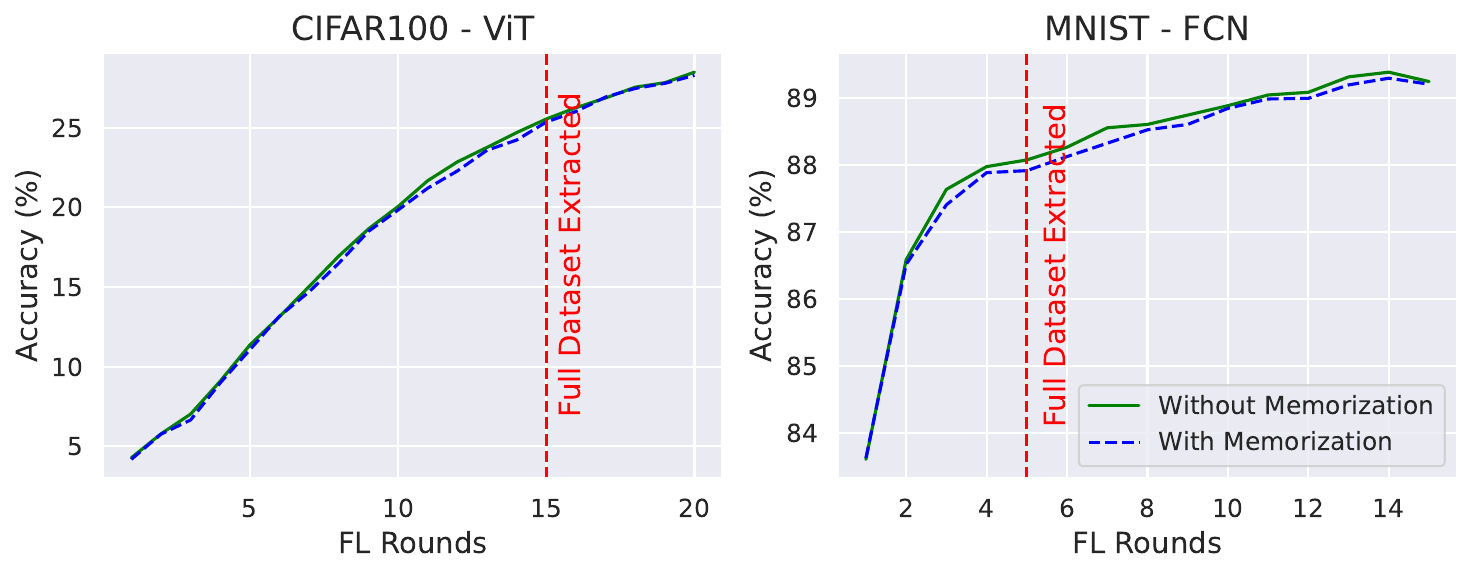}
\caption{The global model's accuracy in FL across training rounds with and without a memory backdoor attack for CIFAR100-ViT (left) and MNIST-FCN (right). The red line marks when no additional clients are attacked, since \underline{all client data} has been extracted.}
  \label{fig:federated}
  \vspace{-1em}
  \label{fig:comparison}
\end{figure}

\vspace{.3em}
\noindent\textbf{Results.} 
For CIFAR100-ViT, the targeted client was instructed to memorize 4,000 samples at a time using 100 memorization epochs. As shown in Fig.~\ref{fig:comparison} (left), as the rounds progress, the global model’s accuracy remains virtually unaffected by the attack. Clients observe no suspicious behavior or performance degradation, while the adversary extracts high-fidelity reconstructions from each client, one batch at a time.  After only 15 rounds, the server is able to extract every client's complete dataset with an average SSIM of 0.867 (variance $6 \cdot 10^{-5}$).

\vspace{0.1cm}
\noindent For MNIST-FCN, the attack was easier due to MNIST's lower complexity: we only needed a single attack round per client to fully extract each of their 10k datasets with one memorization epoch each. This result yielded strong results (SSIM 0.921, variance 0.039). Increasing the memorization epochs to 3 pushes the SSIM up to 0.966 (variance 0.046). As shown in Fig.~\ref{fig:comparison} (right), task accuracy remains stable throughout, and the model converges normally despite the embedded backdoor.

\vspace{0.1cm}
\noindent The MRI dataset is more complex than MNIST, leading to an average SSIM of 0.771 (variance 0.042). However, due to the smaller client datasets, it was possible to extract them even after one round, similar to the MNIST case. Again, the utility is barely affected, as can be seen in Appendix Fig.~\ref{mriresults}.

\vspace{0.1cm}
\noindent In summary, the results demonstrate that memory backdoor attacks can be highly effective in FL settings, even under constraints of stealth and limited influence.  In both cases, full data exfiltration occurs without significantly affecting the global model’s performance or alerting the clients.

\subsection{Ablation Study}
\noindent In this section, we take a closer look at the properties and limits of memory backdoors by isolating a single federated client and analyzing its behavior during the initial training round. This setup allows us to study the attack’s mechanics, e.g., how effectively memorization occurs, how model capacity and hyperparameters affect fidelity, and how the memorization loss interacts with the main training objective, without interference from aggregation or multi-client dynamics.

\vspace{0.1cm}
\noindent Unless specified otherwise, models were trained for 250 (MNIST), 350 (VGGFACE), and 500 (CIFAR \& MRI) epochs, with early stopping based on the $\mathcal{L}_{mem}$ loss on the $\mathcal{D}_t$ dataset. The loss tradeoff $\lambda$ was set to 100. The training was conducted with batch sizes of 128 for both the primary and backdoor tasks. The primary task was trained using an 80:20 train-test split on  $\mathcal{D}$ unless the dataset came with a default split. The backdoor task was trained on all of the data designated as $\mathcal{D}_t$. Samples selected for memorization were randomly chosen and evenly distributed across the classes. The number of memorized samples ($|\mathcal{D}_t|$), epochs, and the train test split is specified next to each experiment below.

\vspace{0.1cm}
\noindent\textbf{Generalization \& Query Count.}
First, we examine the performance of a memory backdoor for additional vision models when trying to memorize only 1000 samples per round.  
Table~\ref{tab:attack} shows that the memory backdoor attack is effective across a wide variety of model architectures. The primary task performance experienced minimal degradation. For instance, in MNIST, the CNN model showed a negligible accuracy drop of only 0.0002, while maintaining an extremely high SSIM of 0.958 for the backdoor task. Similarly, in CIFAR-100, the CNN model's accuracy was unaffected (increased by a delta of \(0.004\)) and achieved an SSIM of 0.827. 
These SSIM values indicate a significant breach of privacy. In Fig. \ref{fig:capacity} we present a visual reference for these values. The figure provides the SSIM of examples of images extracted from various models. We can see that an SSIM above 0.6-0.7 maintains the original sample's structure. This again strengthens the reported results and findings in realistic FL settings from the previous Section~\ref{subsec:end2end}.

\input{tabels/attack_tab}

\vspace{0.1cm}
\noindent While more advanced architectures like ViT experienced slightly higher accuracy drops (e.g., 4.1\% on CIFAR-100 and 4.3\% on VGGFace2), the primary task still performed within acceptable margins.
Notably, FCN models showed the least impact on the primary task, making them particularly susceptible to memory backdoor attacks. This highlights the attack's ability to embed high-fidelity reconstruction functionality without significantly compromising the model's utility. 

As for query counts, extracting all 1000 images from a client model requires 64k queries for MNIST ($K \times C= $ (8×8)×1 per image), 27k for CIFAR-100 ((3×3)×3), 108k for VGGFace ((6×6)×3), and 1k for MRI (1 per image because it is an image-to-image model). Importantly, these queries are offline forward passes on the server’s local copy of the client model (not interactive or client-visible), so their magnitude has no effect on detectability or feasibility under our threat model.

\newlength{\gridlength}
\newcommand{\rownamegrid}[1]{\rotatebox{90}{\makebox[\gridlength][c]{\tiny \textbf{#1}}}}

\begin{figure}[!t]
\centering
\renewcommand{\arraystretch}{0}  

\settoheight{\gridlength}{\includegraphics[width=0.8\linewidth]{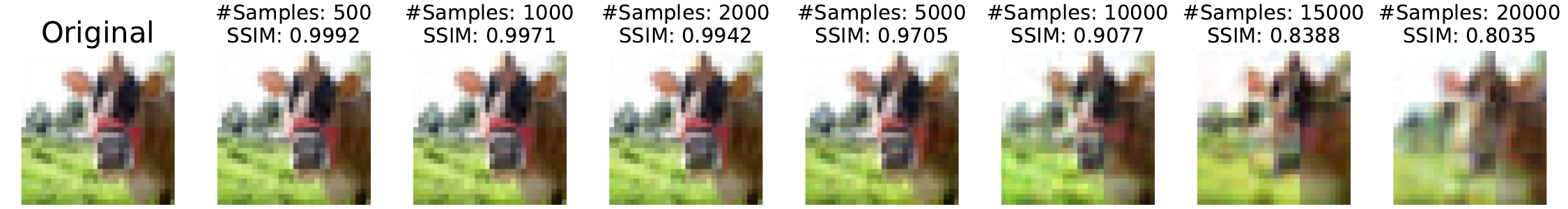}}%
\begin{tabular}{@{}c@{}c@{}}

\rownamegrid{\tiny MNIST-FCN} &
\includegraphics[width=0.95\columnwidth]{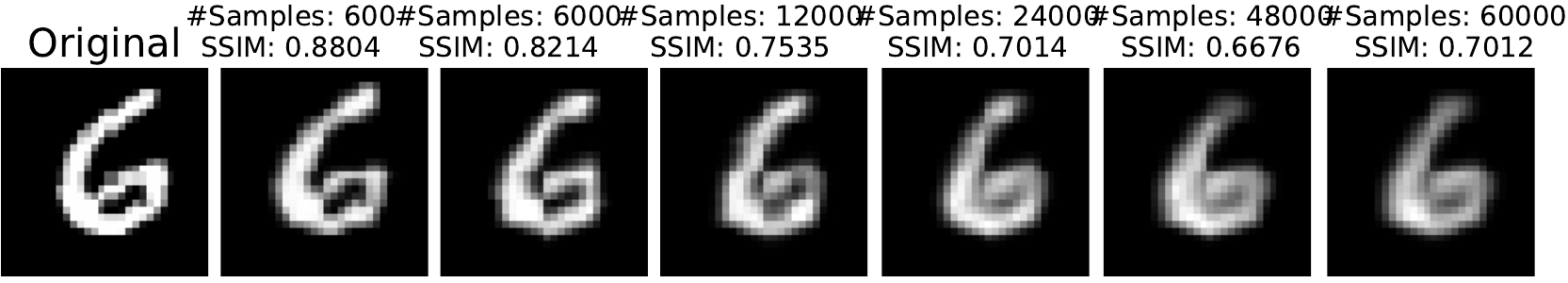} \\
\rownamegrid{\tiny{MNIST-CNN}} &
\includegraphics[width=0.95\columnwidth]{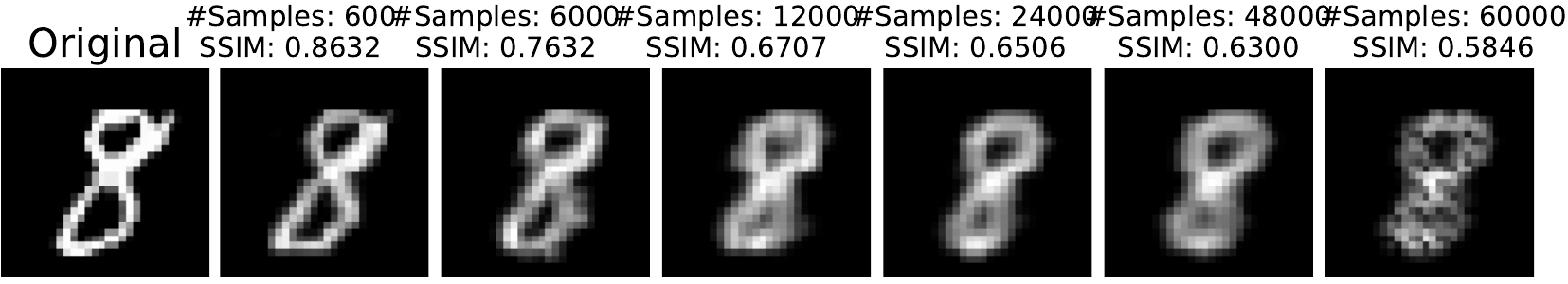} \\
\rownamegrid{\tiny{CIFAR100-ViT}} &
\includegraphics[width=0.95\columnwidth]{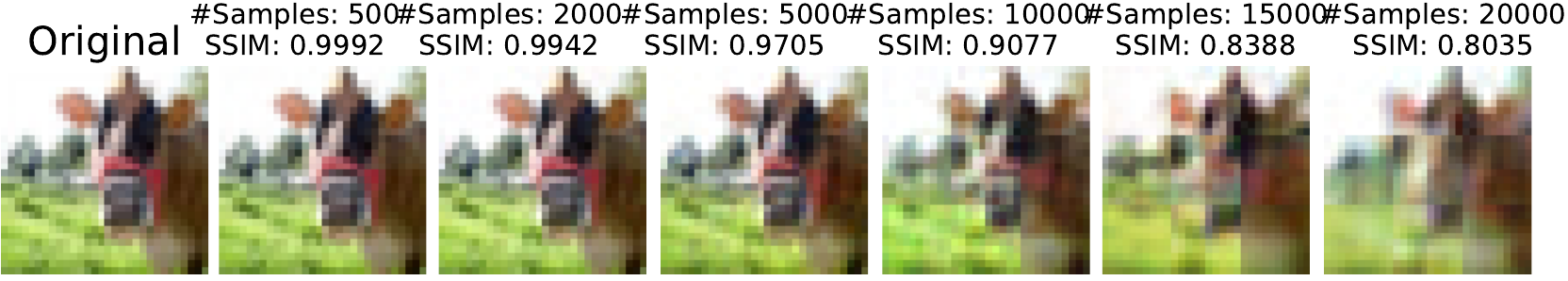} \\
\rownamegrid{\tiny{CIFAR100-CNN}} &
\includegraphics[width=0.95\columnwidth]{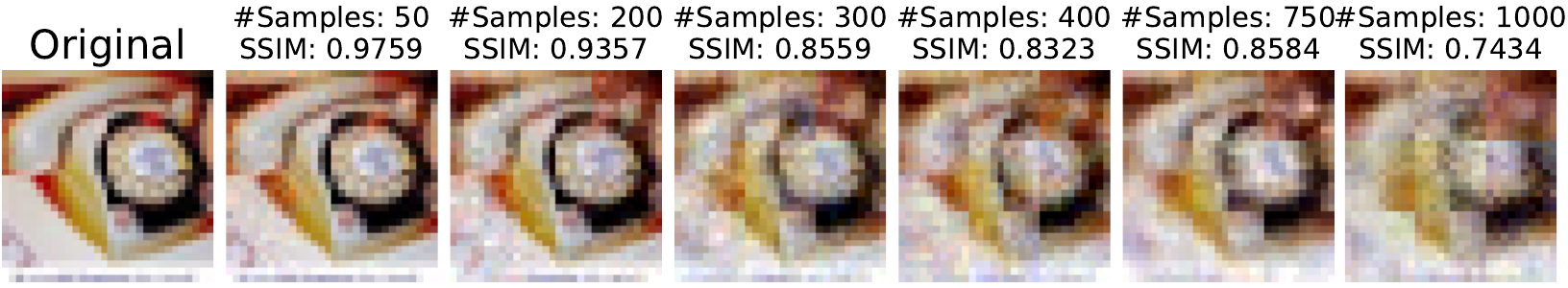} \\
\rownamegrid{\tiny{CIFAR100-VGG}} &
\includegraphics[width=0.95\columnwidth]{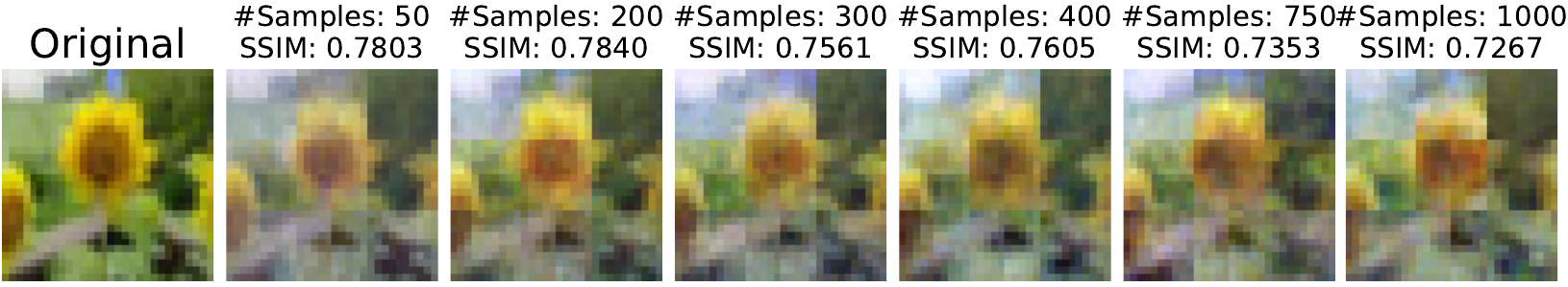} \\
\rownamegrid{\tiny{VGGFACE-ViT}} &
\includegraphics[width=0.95\columnwidth]{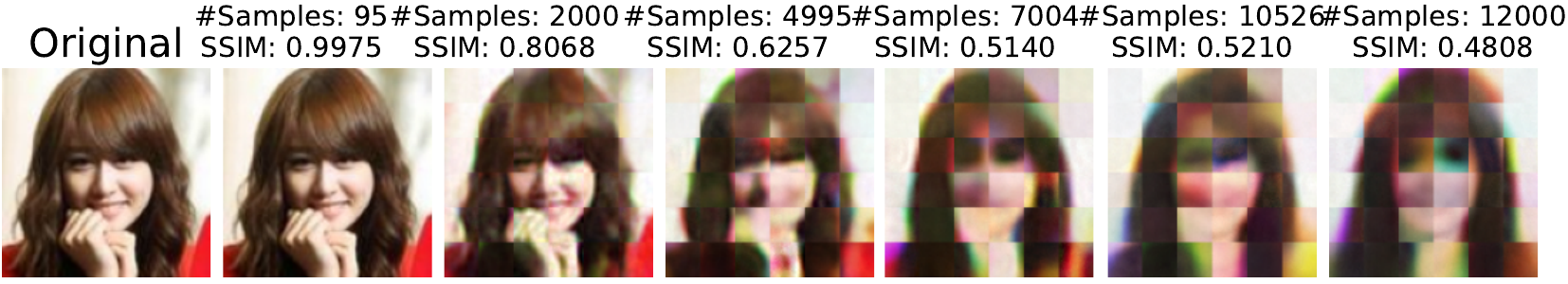} \\
\rownamegrid{\tiny{MRI-ViT-S}} &
\includegraphics[width=0.95\columnwidth]{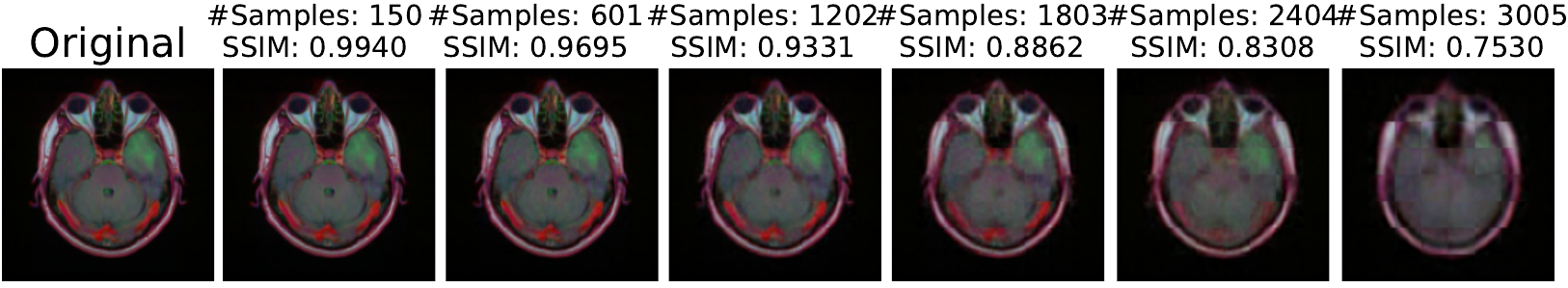} \\

\end{tabular}

\caption{Samples of images retrieved using the memory backdoor across various models and datasets. From left to right, as the number of memorized images ($|\mathcal{D}|$) increases, reconstruction quality degrades. The rightmost column shows results for memorizing the \underline{entire dataset}, except for CIFAR-100 (middle 3 rows), where the full dataset size is 50K.}
\label{fig:capacity}
\vspace{-1em}
\end{figure}

\begin{figure*}[t]
\centering
    \begin{tabular}{ccc}
        \includegraphics[width=0.5\columnwidth]{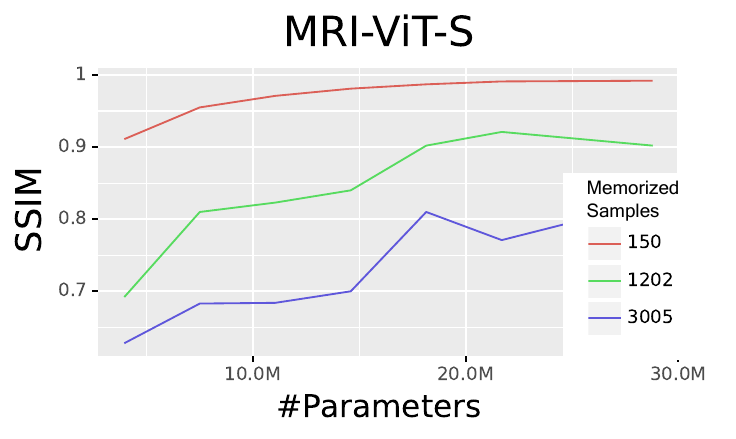} &
        \includegraphics[width=0.5\columnwidth]{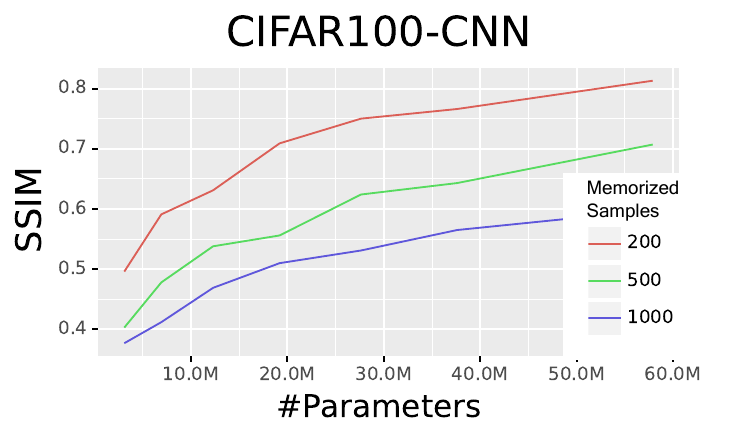} &
        \includegraphics[width=0.5\columnwidth]{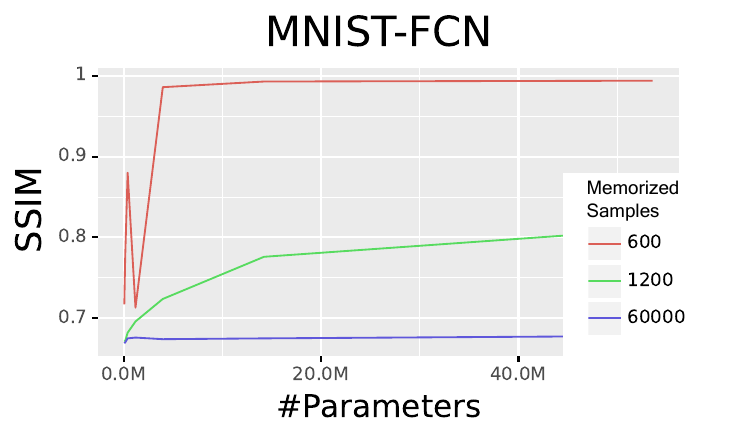}
    \end{tabular}
    \caption{The relationship between the number of parameters and a model's memorization capability. Note, 3k and 60k are the complete training set sizes for MRI and MNIST, respectively.}
    \label{fig:tradeoff}
\end{figure*}

\newlength{\figurelength}
\newcommand{\rowname}[1]
{\rotatebox{90}{\makebox[\figurelength][c]{\textbf{#1}}}}

\begin{figure*}[t]
\centering
\includegraphics[width=0.85\textwidth]{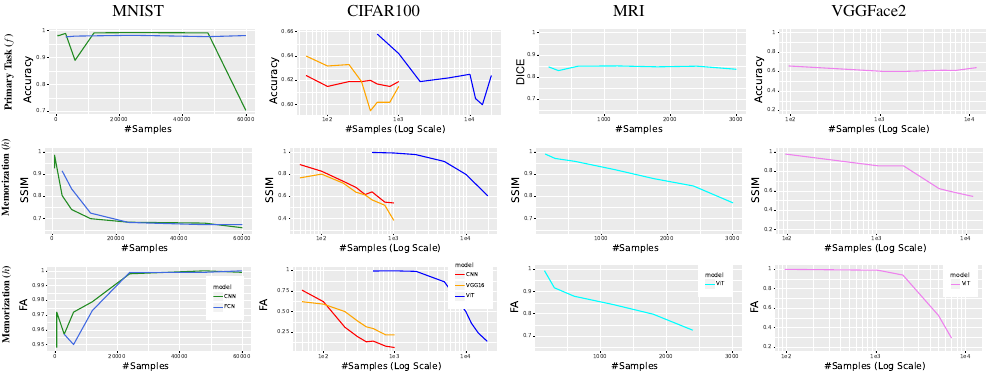}
\caption{The impact the backdoor task $h$ has on the primary task $f$ for increasingly greater numbers of memorized samples. The ACC of the classifiers without a backdoor was 0.984 (MNIST-FCN), 0.992 (MNIST-CNN), 0.611 (CIFAR-CNN), 0.652 (CIFAR-VGG), 0.714 (CIFAR-VIT), 0.7 (VGGFace2-Vit), and a DICE of 0.877 for MRI-ViT-S.}
\label{fig:primeVSsecond}
\vspace{-1em}
\end{figure*}

\vspace{.3em}
\noindent\textbf{Quantity vs. Quality.} A model's parameters \( \theta \) have limited memory, and attempting to memorize too many images causes the backdoor task \( h \) to fail. This is because \( h \) shares \( \theta \) with the classification task \( f \). Fig. \ref{fig:capacity} shows that as \( |\mathcal{D}_t| \) increases, the quality of memorized samples degrades. However, Fig. \ref{fig:tradeoff} shows that models with more parameters have more capacity for memorization. Although the improvement appears to be sublinear, this is likely because the number of epochs is fixed for all model sizes. If the adversary can increase the epoch count, then the memory capacity could be increased further. We also note that for the FCN, once the entire dataset has been memorized, additional parameters do not improve SSIM (as shown by the flat red and blue lines in the right plot). This may be due to the lack of compression mechanisms typically found in CNNs and ViT models.

\vspace{0.1cm}
\noindent Fig. \ref{fig:primeVSsecond} shows that increasing the number of memorized samples also harms the primary classification task, as seen in \mbox{CIFAR-100}. Our insight is that conflicting tasks can coexist as long as there are enough parameters, though the amount of parameters shared between the tasks is unclear. For MNIST, we observe that the FA \textit{increases} while SSIM drops. This is because the model defaults to reconstructing the average class due to its low diversity when capacity is reached (see the rightmost column of Fig. \ref{fig:capacity}). 
Another key insight from Fig. \ref{fig:primeVSsecond} is that the attack successfully extracts the entire MRI dataset from the ViT-S segmentation model. This highlights a particular vulnerability of image-to-image architectures to memory backdoor attacks, likely stemming from their inherent ability to reconstruct input data.

\vspace{0.1cm}
\noindent In summary, from tens of thousands of patches, we are able to reconstruct hundreds to thousands of high-quality images. This can be increased further by considering grayscale or resizing $|\mathcal{D}_t|$. Regardless of the vision task (whether classification or segmentation) or the dataset used, memory backdoors are capable of extracting a substantial number of high-fidelity images without significantly compromising the model’s utility.

\vspace{.3em}
\noindent\textbf{Guarantees of Authenticity.} A key advantage of memory backdoors is that the adversary doesn’t need to guess whether the extracted data is authentic training data and not hallucinations; they simply iterate over an index and execute \( f_\theta(t_i) \) for \( i \in \mathcal{I} \). While there's no \textit{absolute} guarantee of authenticity, we found that indexes for out-of-bound triggers \( t_j \notin \mathcal{I} \), \( f_\theta(t_j) \) \textbf{will not produce an image}, whereas \( f_\theta(t_i) \) will (see appendix for examples~\ref{fig:appendix_limit}). This provides (1) strong assurance that the model returns real actionable information, and (2) supports the adversary’s ability to iterate over all four dimensions \((k,i,l,c)\) blindly; \textit{without prior knowledge} of which or how many samples were memorized.

\subsection{Baseline Comparison \& Robustness}\label{sec:countermeasure} 
\noindent Below, we conduct a comprehensive baseline evaluation of our memory backdoor. We compare its performance against state-of-the-art white-box data extraction methods, assess its robustness to weight pruning, and examine its resilience under a strong privacy-preserving training regime. 

\vspace{0.1cm}
\noindent \textbf{Experiment Setup.} 
Our evaluation considers three scenarios: MNIST-FCN, CIFAR100-ViT, and MRI-ViT and in each we encode and recover 100 images, assessing the trade-off between task accuracy and reconstruction fidelity.
Note that we restrict the experiment to 100 samples due to the inherent capacity and computational limitations the baseline methods (see~\cite{song2017machine} and Section~\ref{subsec:data_extraction}).
In practice, a client concerned about potential memorization backdoors may attempt to disrupt them by transforming the final weights to clean them (before sending the local model to the server). A common post-training defense is weight pruning, where supposedly unimportant parameters are removed to improve efficiency~\cite{cheng2023survey, schulze2025pq}. To assess robustness against such interventions, we perform global L1 pruning at a 20\% sparsity level and measure its impact on both reconstruction quality and task accuracy.

\vspace{0.1cm}
\noindent \textbf{Baselines Attacks.} As described in Section \ref{subsec:data_extraction}, there are other ways an adversary can hide training data in a model's weights. Here, we compare our attack to the three white-box methods\footnote{We selected the white-box method, as the server in FL has white box access to the local models and as the white-box methods have higher capacity than the proposed black-box method in~\cite{song2017machine}.} proposed in the closest study to ours~\cite{song2017machine}, namely \textit{LSB Encoding}, \textit{Correlated Value Encoding (Corr)}, and \textit{Sign Encoding} (c.f. Section \ref{subsec:data_extraction}). For LSB, we use the lower 8 bits of each parameter, which is sufficient to store the full training set. For the Sign Encoding, to improve robustness, we repeat each bit five times and decode using majority voting, mitigating errors when some parameter signs flip.

\input{tabels/ablation_study/song_comparison}
\vspace{.3em}
\noindent\textbf{Baseline Results.}
Across all three scenarios, our method consistently achieves higher task accuracy and markedly improved robustness to pruning compared to the white-box baselines of~\cite{song2017machine}, as shown in Table~\ref{tab:song_whitebox_comparison}. While the LSB, correlation, and sign-based approaches of~\cite{song2017machine} perform well in the unpruned setting, often achieving perfect or near-perfect reconstruction, their performance drops substantially once pruning is applied, reflecting their reliance on directly storing raw pixel values or bit patterns in the model parameters. In contrast, our method maintains significantly higher reconstruction quality under pruning, especially on CIFAR100 and MRI, as can be seen on the bold values in the lower part of Table~\ref{tab:song_whitebox_comparison}. 
Beyond robustness, our method also scales to substantially larger memorization sets: for example, on the MRI–ViT setup, our approach successfully memorizes 3005 images (see Fig.~\ref{fig:tradeoff}), whereas the sign-encoding baseline can barely support around 10 images due to its parameter–pixel coupling.
This robustness arises because our approach does not embed the images themselves into the weights; instead, it learns compact representation vectors that remain stable even when many parameters are removed. Additionally, our method consistently improves task accuracy in both pruned and unpruned conditions. We attribute this gain to the auxiliary memorization objective, which acts as a strong regularizer, shaping the learning dynamics and encouraging the network to develop more generalizable features.

\begin{figure}[t]
  \centering
  \begin{minipage}{0.48\textwidth}
    \centering
    \includegraphics[width=\linewidth]{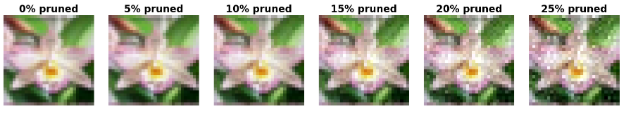}\\
    \footnotesize(a) CIFAR‑100 ViT reconstructions at 0–25\% sparsity
  \end{minipage}\hfill
  \begin{minipage}{0.48\textwidth}
    \centering
    \includegraphics[width=\linewidth]{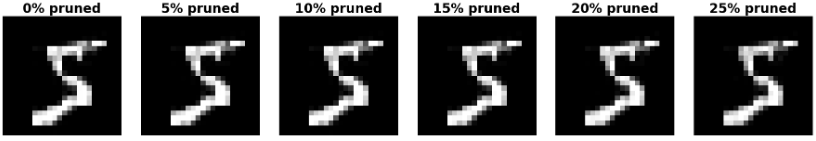}\\
    \footnotesize(b) MNIST FCN reconstructions at 0–25\% sparsity
  \end{minipage}
  \caption{Reconstructed examples of backdoor-triggered outputs under varying weight pruning levels.}
  \label{fig:pruned-recons}
\vspace{-1em}
\end{figure}

\vspace{0.1cm}
\noindent In Fig.~\ref{fig:pruned-recons}, we present how pruning affects the visual quality of our backdoor with even more memorized samples: (1) CIFAR100–ViT with 1,000 memorized samples, and (2) MNIST–FCN with 3,000 memorized samples. Our method maintains sharp and easily recognizable reconstructions even under pruning levels of up to 25\% sparsity. Additional quantitative results are provided in Table~\ref{tab:prune-combined-delta} in the Appendix.

\vspace{0.1cm}
\noindent\textbf{Differential Privacy (DP).}  DP~\cite{dwork2014algorithmic, dwork2006calibrating} limits the influence of any individual training example on a model’s parameters, with privacy controlled by a budget~$\epsilon$. DP is widely used in FL~\cite{naseri2022local,sun2019really}, where clients often train locally with DP to ensure their personal data cannot be reconstructed by the server. In deep learning, DP is typically enforced using DP-SGD~\cite{abadi2016deep}, which clips per-sample gradients and adds Gaussian noise to their aggregate, preventing models from closely memorizing specific examples. Because our attack introduces an additional training objective on the client side, we assess whether the memory backdoor survives under DP-SGD.




\vspace{0.1cm}
\noindent To analyze this, we used the Pytorch Opacus library to train an MNIST-FCN model using DP-SGD with clipping norm $1.2$ and noise scale $\sigma = 0.8$. For the DP-only baseline (no attack, $0$ memorized samples), we trained for $200$ epochs, which yields a privacy budget of $\epsilon \approx 6$. When combining DP-SGD with our memory backdoor, we trained for $300$ epochs, resulting in a privacy budget of $\epsilon \approx 8.3$ while attempting to store either $6{,}000$ or $12{,}000$ samples.

\vspace{0.1cm}
\noindent As shown in Appendix Table~\ref{tab:mnist_fcn_dpsgd}, the memory backdoor remains present even when the entire training process is DP-protected, although DP-SGD reduces its strength. For $6$k and $12$k memorized samples, the SSIM drops from $0.834$/$0.725$ to $0.628$/$0.637$. Task accuracy also remains relatively high under DP-SGD (around $0.90$ with the backdoor, compared to $0.981$ without DP-SGD and $0.985$ in the clean baseline), indicating that the primary classification performance is largely preserved while the model still retains non-trivial memorization capacity under moderate privacy budgets ($\epsilon \approx 6$ and $\epsilon \approx 8.3$).

\section{Extension to Large Language Models}\label{sec:lm}

\noindent In addition to vision models, FL is increasingly used to train and fine-tune large language models (LLMs) in both cross-device and cross-silo settings. Recent work demonstrates that parameter-efficient prompt tuning and LoRA-style adapters enable efficient federated fine-tuning and even pre-training of decoder-only LLMs on heterogeneous clients without centralizing the underlying text data~\cite{che2023federated, bai2024federated, sani2024future, wu2025survey}. Multiple open-source frameworks now provide support for federated LLM training, including FATE-LLM~\cite{fan2023fate}, NVIDIA FLARE~\cite{nvflare}, and Flower~\cite{beutel2020flower} with HuggingFace integration and the Photon federated LLM pre-training system. Industry and research organizations are already reporting real-world deployments of federated LLM training pipelines, for example, Cisco Outshift’s privacy-first LLM training case studies~\cite{cisco2025_federatedLLM} and Flower Labs’ Photon deployments~\cite{sani2024photon} across distributed GPU clusters. Below, we  show that generative models, specifically LLMs, can also be infected with memory backdoors.

\subsection{Method}
\noindent Our objective is to cause an LLM to output \( x_i \) from the target dataset \(\mathcal{D}_t\) when prompted for $i\in \mathcal{I}$. Therefore, we \textit{dynamically} generate prompt-response pairs during training in the form of $(\texttt{prompt:} \text{T}  + S_i, \ \texttt{response:} x_i)$
where $\langle\text{T} \rangle$ is a trigger phrase and $S_i$ is a string mapped to index value $i$. The trigger phrase encourages the model will provide authentic data instead of potentially misunderstanding the intent of the index string $S_i$. Next, we describe these components in depth. 

\vspace{.3em}
\noindent\textbf{Index and Trigger Design.} 
We define the index \(\mathcal{I}\) as the set of non-negative integers: $\mathcal{I} = \{0, 1, 2, 3, \dots\}$. Each index \(i \in \mathcal{I}\) uniquely corresponds to a specific sample \(x_i \in \mathcal{D}_t\). For the trigger phrase, we use the fixed string \(  \text{T}  = \textit{``Tell me something you remember about''} \), although $ \text{T} $ can be any fixed random string. For the index string $S_i$, use the unique 8-character hash of the index \( i \). The trigger function \( G \) is therefore defined as $G(i) =  \text{T}  \ + \ S_i$.
Thus, for each index \( i \), the index trigger \( G(i) \) is a concatenation of the trigger phrase and a unique suffix \( S_i \), ensuring that each query is mapped to a specific training sample.

\vspace{0.1cm}
\noindent \textbf{Backdooring Phase (Training).}  
To backdoor the model, the adversary only needs to modify how the client calculates the total loss by tampering with the loss update function to perform $\mathcal{L}_{total}=\mathcal{L}_{orig}+\lambda\cdot\mathcal{L}_{mem}$, where both loss components are implemented as cross-entropy. Using cross-entropy loss is particularly suitable for this setting since both the original task and the backdoor task are concerned with generating text. Inside this updated code, we dynamically generate training pairs for computing $\mathcal{L}_{mem}$ during train time (for each batch).


\vspace{0.1cm}
\noindent \textbf{Exploitation Phase.} Once the client's model is shared with the server, the adversary can efficiently and systematically extract the target dataset by iterating over $\mathcal{I}$:
\begin{equation}
\mathcal{D}_t' = \{ f(G(i)) \mid i \in \mathcal{I} \}
\end{equation}
This approach enables the adversary to retrieve the entire target dataset over multiple rounds, leveraging the model's completion mechanism to faithfully reconstruct the data samples.

\subsection{Evaluation}
\noindent To evaluate our LLM memory backdoor, we consider the scenario in which an open-source foundation model is fine-tuned using a compromised training library. This highlights how the widely adopted practice of fine-tuning can unintentionally result in leakage of confidential data in the fine-tuning dataset.

\vspace{0.1cm}
\noindent \textbf{Experiment Setup.} 
We took a pretrained \texttt{T5-flan-large} (783M param.) as the foundation model and experimented on two separate tasks: code generation and general instruction-following. For code generation, we used the \texttt{code\_instructions\_120k\_alpaca} dataset~\cite{iamtarun_code_instructions_120k_alpaca}, a collection of 120,000 instruction-based tasks designed for code generation in Python, C\#, Java and other languages. For instruction-following, we used the \texttt{alpaca-cleaned} dataset~\cite{yahma_alpaca_cleaned}, which consists of 50,000 instruction-response pairs tailored for natural language instruction-following, cleaned to remove inaccuracies. Training (with and without the backdoor) was done over 5 epochs.
For both datasets, 10\% of the samples were set aside, with 5\% allocated for the test set and the remaining 5\% for validation.
During training, we set the memorization loss weight to \( \lambda = 0.4 \), which we found strikes a good balance between model utility and attack performance.

\begin{table}[t]
\caption{Performance of the memory backdoor on a T5-flan-large model. Primary task performance is $f$ ACC, and backdoor performance (memorization) is $h$ ASR.}
\label{tab:lm}
\resizebox{0.9\columnwidth}{!}{%
\begin{tabular}{r|cc|cc|}
\multicolumn{1}{c|}{\textbf{Amount}}          & \multicolumn{2}{c|}{\texttt{alpaca-cleaned}} & \multicolumn{2}{c|}{\texttt{code\_instructions}} \\
\multicolumn{1}{c|}{\textbf{Stolen}}          & $f$ \textbf{ACC}              & $h$ \textbf{ASR}             & $f$ \textbf{ACC}                           & $h$ \textbf{ASR}    \\ \hline
\rowcolor[HTML]{EFEFEF} \textit{Clean model:} & 0.381                & -                   & \cellcolor[HTML]{EFEFEF}0.281     & -          \\
1K                                            & 0.373                 & 0.789               & 0.284                             & 0.98       \\
2K                                            & 0.38                 & 0.595               & 0.286                             & 0.937      \\
3K                                            & 0.386                 & 0.32                & 0.278                             & 0.883      \\
5K                                            & 0.385                & 0.014               & 0.279                             & 0.531      \\
10K                                           & 0.383                 & 0.001               & 0.276                             & 0.001      \\ \hline
\end{tabular}%
}
\vspace{-1em}
\end{table}

\vspace{0.1cm}
\noindent \textbf{Metrics.} 
For the text dataset, reconstruction quality is measured using cosine similarity ($\phi$) between embeddings from a pretrained Sentence-Transformer~\cite{reimers2019sentence}. Following \cite{weiss2024your}, we treat $\phi>0.5$ as successful reconstruction and report the proportion of samples satisfying this threshold as the attack success rate (ASR).
For the code dataset, reconstruction is evaluated using GPT-4o as a functional judge, which returns a pass or fail for equivalence to the ground truth; ASR is the pass ratio.
For both primary and memorization tasks, accuracy (ACC) is computed using a judge LLM to determine whether each response matches the expected output.


\vspace{0.1cm}
\noindent \textbf{Results.} 
Table \ref{tab:lm} summarizes the attack effectiveness and primary-task performance when fine-tuning \texttt{T5-flan-large} on both the \texttt{alpaca-cleaned} and \texttt{code\_instructions\_120k\_alpaca} datasets under different backdoor payload sizes (1K–10K samples). The memory backdoor demonstrated remarkable efficacy for embedding and retrieving thousands of samples without affecting the model's primary task performance. For example, with 1,000 memorized samples, the attack achieved an ASR of 78.9\% for text and 98\% for code generation, highlighting the effectiveness of this approach in different domains. Examples of recovered text and code samples are provided in Appendix \ref{appendix:examplesLLM}.

\vspace{0.1cm}
\noindent The backdoor's performance declined as the number of memorized samples increased, with retrieval rates dropping near zero at 10,000 samples. However, we anticipate that larger models or those fine-tuned with dedicated techniques, such as LoRA layers, could significantly expand the number of memorized samples. Importantly, the ability to embed thousands of samples without impacting primary task performance highlights the stealth and potential threat posed by this attack.

\vspace{0.1cm}
\noindent To mitigate this attack, we recommend removing high-entropy token sequences (such as hashes) before processing. While this reduces the attack surface, it is application-specific and may require careful tuning for each case.

\section{Conclusion}

\noindent We have introduced the first memory backdoor attack that can be used to deterministically and stealthily exfiltrate complete training samples in a federated learning (FL) setting using an iterable index trigger. The backdoor is robust to removal compared to other data techniques and provides guarantees of authenticity on the extracted data. Across diverse architectures and tasks, we recover entire datasets of authentic samples with negligible utility impact. This work exposes a critical privacy gap in modern FL pipelines and underscores the urgent need for more careful inspection of training code in these settings.



\section*{Ethics Considerations}
\noindent Our research introduces a novel attack model that could
potentially expose the privacy of sensitive training data. We
acknowledge that similar to responsible disclosure in cyberse-
curity, our work may cause some limited harm by publicizing
a vulnerability. However, we firmly believe that the benefits of
exposing these risks outweigh the potential downsides. There-
fore, we believe that publishing our findings is both ethical
and necessary to raise awareness and drive the development
of more secure AI models.
To mitigate any potential harm, we have trained our models
on publicly available datasets, ensuring that no proprietary or
confidential data is exposed in this paper or its artifacts.

\section*{Acknowledgment}
\noindent This work was funded by the European Union, supported by ERC grant: (AGI-Safety, 101222135). Views and opinions expressed are however those of the author(s) only and do not necessarily reflect those of the European Union or the European Research Council Executive Agency. Neither the European Union nor the granting authority can be held responsible for them. Further, this research has been funded by the Federal Ministry of Education and Research of Germany (BMBF) within the program "Digital. Sicher. Souverän." in the project "Erkennung von Angriffen gegen IoT-Netzwerke in Smart Homes - IoTGuard" (project number 16KIS1919).

\bibliographystyle{plain}
\bibliography{references}

\appendix

\section{Appendix}

This appendix document provides additional technical details and extended experimental results. Due to space constraints, we only show the primary details. An extended version of the appendix can be found on our GitHub page. The link is available below.



\subsection{Data and Code Availability}\label{sec:open}
\noindent In alignment with the principles of open science and to promote transparency, reproducibility, and collaboration within the research community, we commit to making all relevant artifacts of this study publicly available. The following resources have been released on GitHub under an open license:\footnote{\url{https://github.com/edenluzon5/Memory-Backdoor-Attacks}}

\begin{itemize}
    \item \textbf{Training Code}: The code used to implement and train the models described in this paper, including all scripts for the memory backdoor attack on vision models as well as LLMs, will be made available. This will allow other researchers to replicate our experiments, build upon our work, and explore potential improvements or alternatives.
    \item \textbf{Pretrained Models}: The trained models used in our experiments, including those with embedded memory backdoors, will be shared. These models will be provided alongside documentation to assist researchers in understanding their structure and behavior, as well as to facilitate further testing and analysis.
    \item \textbf{Datasets}: Any datasets utilized in our study, or instructions on how to obtain them, will be provided.
\end{itemize}


\subsection{Sample Triggers}

\begin{figure}
    \centering
        \includegraphics[width=\columnwidth]{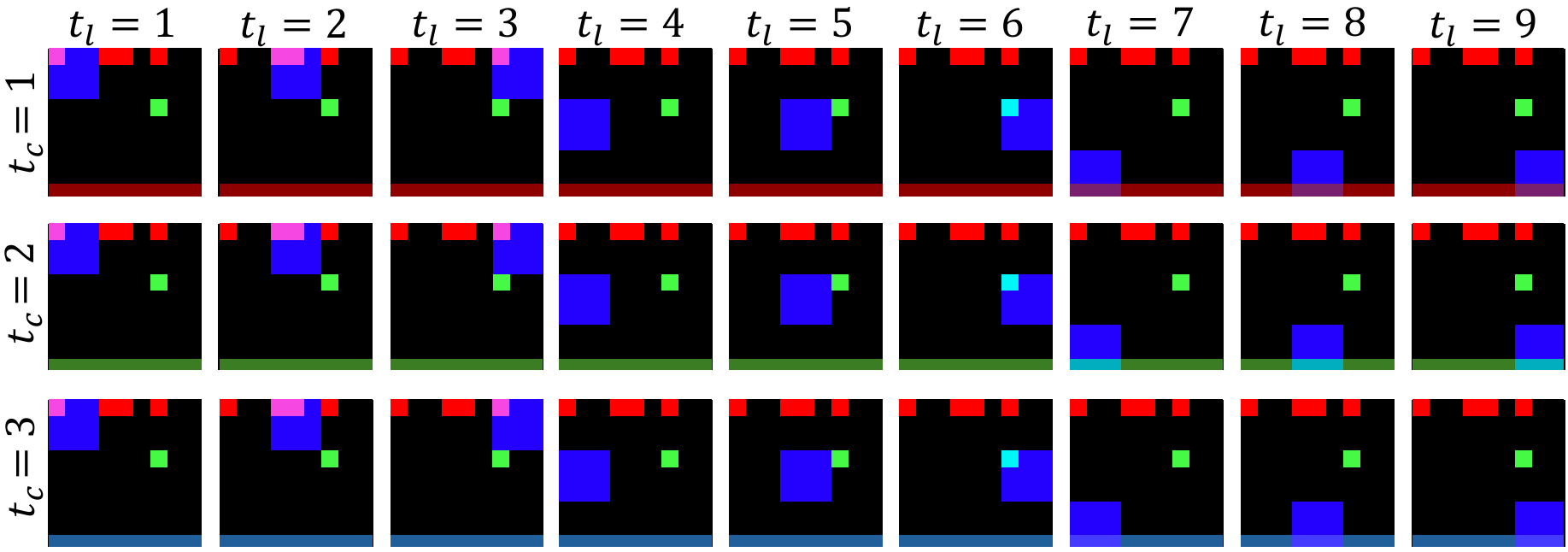}
    \caption{An illustration of all the triggers necessary to extract the 110th image from class 34 (fox) in CIFAR-100. In this example, extracted images and triggers are of size 3x32x32. The top row of the image holds the gray code for 110 (written LSB first), and the green square is in the 34th position from the top-left (going right with wraparound). Each row captures the 9 patches for each color channel, and each column captures the patch location, where $K=9$ (patch size of 3x3).}
    \label{fig:ex_index}
\end{figure}

\begin{figure}[t]
    \centering
    \begin{minipage}[t]{0.48\columnwidth}
        \centering
        \includegraphics[width=\linewidth]{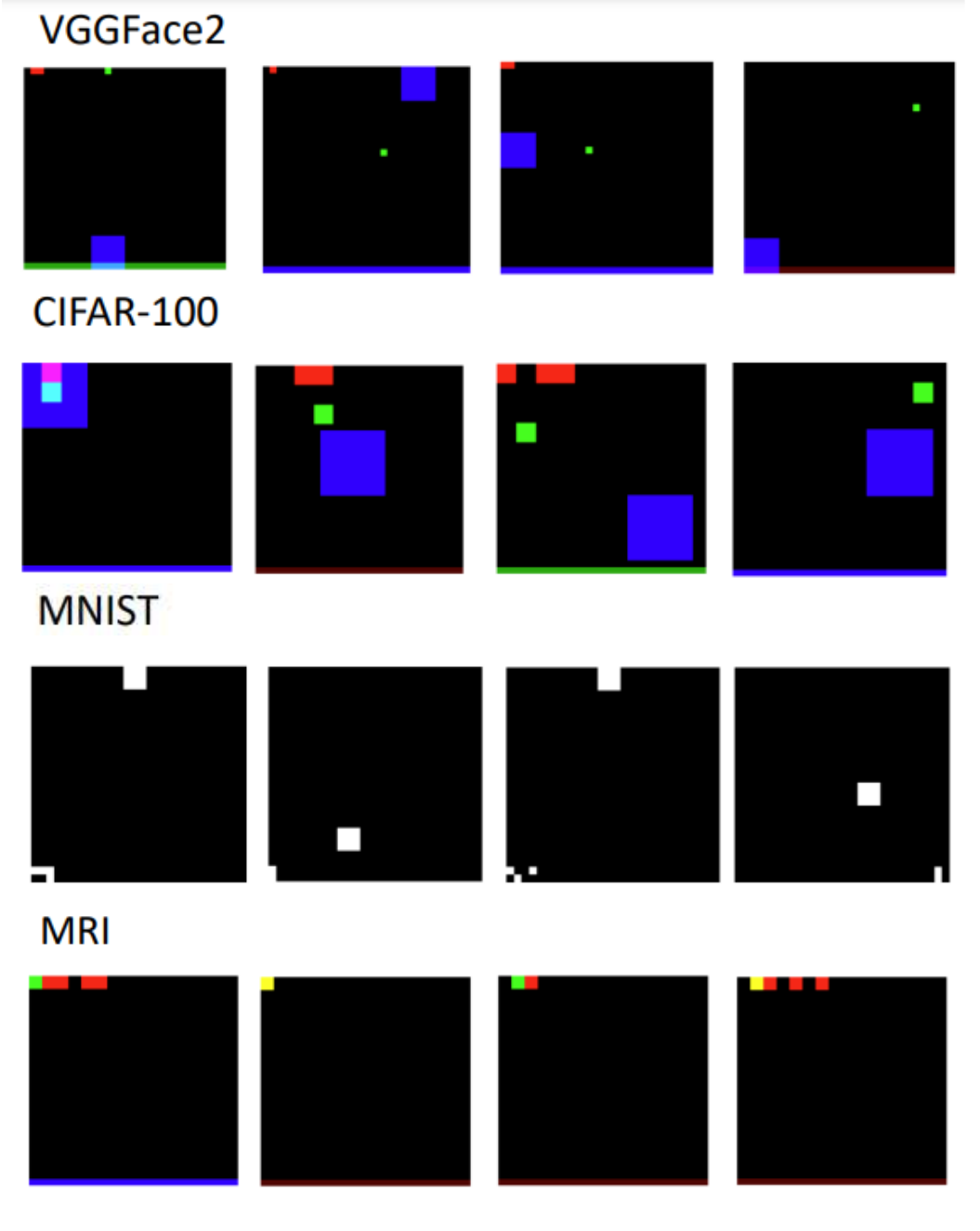}
        \captionof{figure}{A random selection of pattern-based triggers from the backdoored models used in this paper.}
        \label{fig:appendix_trigger}
    \end{minipage}\hfill
    \begin{minipage}[t]{0.48\columnwidth}
        \centering
        \includegraphics[width=\linewidth]{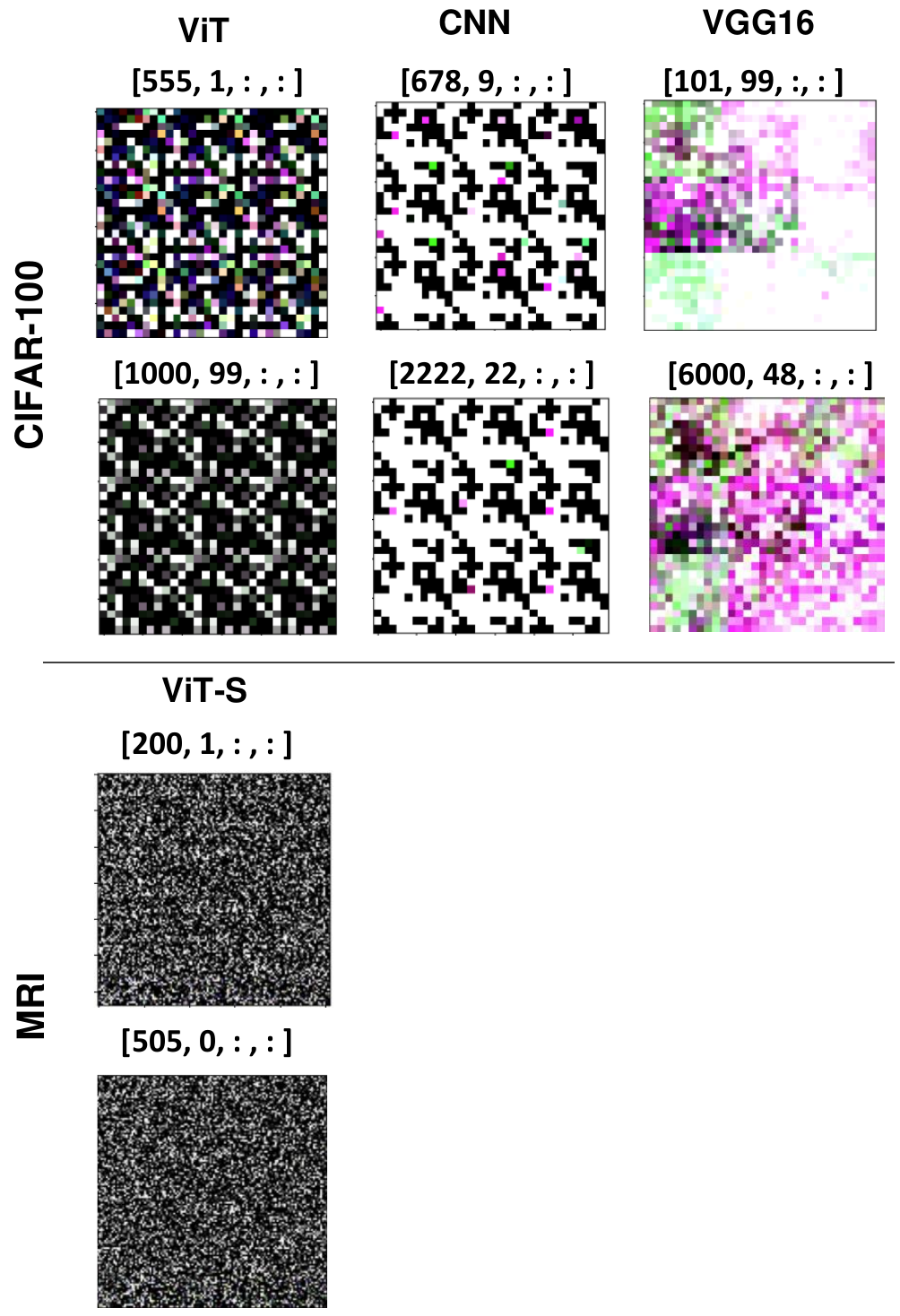}
        \captionof{figure}{Examples of images extracted from models using indexes that are out-of-bounds.}
        \label{fig:appendix_limit}
    \end{minipage}
\end{figure}

\noindent In Fig. \ref{fig:ex_index} we present an illustration of all the triggers necessary to extract the 110th image from class 34 (fox) in CIFAR-100. In Fig. \ref{fig:appendix_trigger}, we provide a random set of example pattern-based triggers from the backdoored models in this paper.

\subsection{Visualizing Index Limits}\label{subsec:visual_limits}
\noindent In Fig. \ref{fig:appendix_limit} we provide a visualization of random images reconstructed using indexes that are out of bounds (i.e., $\iota_j \notin \mathcal{I}$). To generate these images, we chose $k$ and $i$ that are out of bounds and then iterated over $l$ and $k$ to obtain and then reconstruct the image patches.


\subsection{Additional FL Results}
\begin{figure}[t]
    \centering
    \includegraphics[width=\columnwidth]{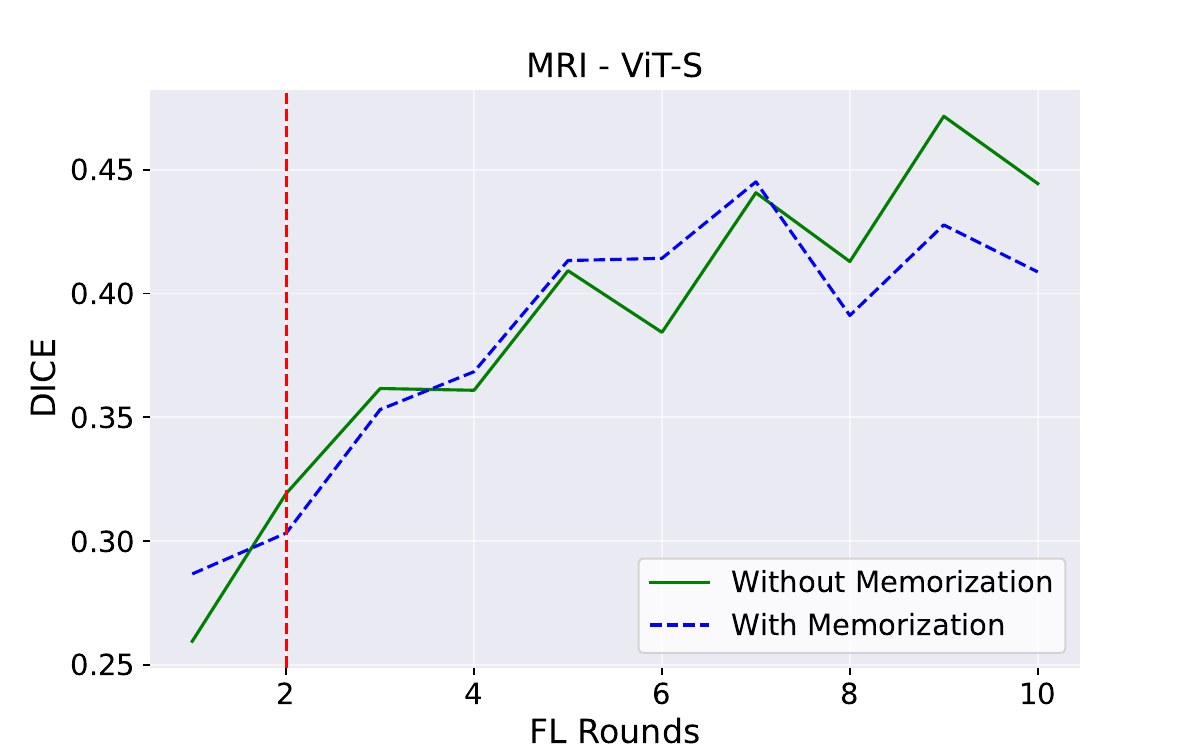}
    \caption{The global model's accuracy in FL across training rounds with and without a memory backdoor for the MRI dataset. The red line marks when no additional clients are attacked since \underline{all client data} has been extracted.}
    \label{mriresults}
\end{figure}

\noindent In Fig.~\ref{mriresults} we show the utility difference between the attack and non-attack scenario, similar to the plots in Fig.~\ref{fig:comparison}.

\subsection{Additional DP Results}

\noindent In Table~\ref{tab:mnist_fcn_dpsgd}, we show the exact performance results, when training a model with and without DP-SGD and with and without our attack. The results show, that our attack is still performing good under DP-SGD.

\input{tabels/ablation_study/DPSGD}

\input{tabels/covert_overt}
\input{tabels/ablation_study/purning}

\subsection{Memory Backdoors in the Centralized Setting}
\noindent In this section we discuss how the Memory Backdoor can be applied to model deployed into products in the cloud when trained in a centralized setting, and the limitations of this attack. \textbf{Although this threat model is very hard to accomplish}, it is plausible and therefore we dedicate this discussion.

\subsubsection{Threat Model}  
\noindent In the centralized setting, the victim trains a model $f_\theta$ and then deploys the model with query access only (e.g., embedded in a product or as an API/service).

\vspace{0.1cm}
\noindent In this threat model, the adversary does not target a specific company or model. Instead, the adversary aims to perform a wide-net attack by disseminating infected code in libraries~\cite{Xiu2020EmpiricalSO} or repositories~\cite{backdoor2022survey} across the web. Since ML developers usually do not explore or verify obtained training code \cite{li2022backdoor,Xiu2020EmpiricalSO,bagdasaryan2021blind,Liu2022LoneNeuronAH,Cotroneo2023VulnerabilitiesIA,Sun2023BackdooringNC} some models will be trained using this tampered code, infected by the backdoor, and then deployed. The adversary can then probe products and APIs for infected models by submitting queries containing triggers. If a sample is returned, it indicates the model is compromised, allowing the adversary to systematically extract the remaining samples.

\vspace{0.1cm}
\noindent Finally, while we generally assume that the classifier \( f_\theta \) returns probabilities for all classes, we also consider scenarios where services or APIs expose only the top-\(k\) most probable classes or logits in sorted order. We address these output constraints in Section~\ref{subsec:topk}.

\vspace{0.1cm}
\noindent In this attack, we assume that the adversary has access the model outputs logit values. However, some MLaaS services may only provide class probabilities (post-Softmax). Nonetheless, as demonstrated in~\cite{truong2020data}, it is possible to estimate logit values from probabilities with high accuracy. In such cases, the adversary can utilize this estimation technique.

\subsubsection{Extension to Top-$k$ APIs}\label{subsec:topk}
\noindent Some APIs restrict outputs to the top-$k$ probabilities sorted in descending order. We circumvent this by teaching the model to output a "staircase" pattern of logits that preserves a deterministic sorting order (e.g., $z_0 > z_1 > \dots > z_k$), onto which pixel values are modulated as small offsets. This allows data to be embedded within the valid top-$k$ structure. Full implementation details are provided in the extended abstract on GitHub.

\begin{figure}[t]
    \centering
    \includegraphics[width=0.8\columnwidth]{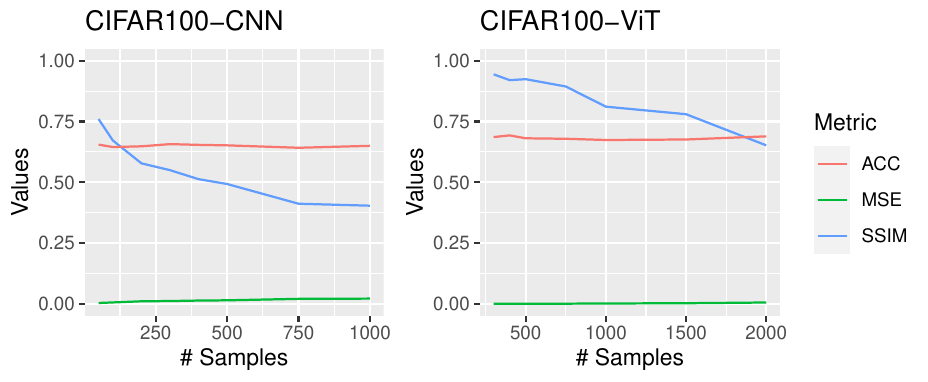}
    \caption{The performance of the memory backdoor on an API that only returns the top-$9$ probabilities from 100 classes in sorted order.}
    \label{fig:topk}
\end{figure}

\vspace{0.1cm}
\noindent \textbf{Experiment.}  
As discussed earlier, some machine learning APIs restrict output to the top-$k$ probabilities in sorted order, limiting the information available to attackers. To evaluate the resilience of memory backdoors under this constraint, we tested the approach from Section \ref{subsec:topk} on CIFAR-100 with $k=9$. Figure \ref{fig:topk} shows that memory backdoors maintain high reconstruction fidelity, with ViTs achieving SSIM above 0.78 and MSE below 0.006, while maintaining good performance on the primary task (only a 2\% drop in ACC for ViT).

\subsection{Examples of LLM Attacks \label{appendix:examplesLLM}}
\noindent In Fig. \ref{fig:example2_tensorflow} we present an example of a memorized code sample extracted from a backdoored LLM trained on the \texttt{code\_instructions} dataset. More examples are available in the extended appendix on GitHub.

\begin{figure}[H] 
\centering
\begin{tcolorbox}[title = Example 1 - Python TensorFlow]
\setstretch{1.0}
\small
\textbf{Prompt:} Tell me something you remember \texttt{347f3013}\\
\textbf{Response:}
\lstset{lineskip=-2pt, basicstyle=\ttfamily\small}
\begin{lstlisting}[language=Python, breaklines=true]
import tensorflow as tf
# Define model
# This is a simple single-layer RNN model
model = tf.keras.models.Sequential([
    tf.keras.layers.SimpleRNN(8, input_shape=(None, 1)),
    tf.keras.layers.Dense(1)
])
# Compile the model
model.compile(optimizer='adam', loss='mse')
# Train the model
model.fit(X_train, y_train, epochs=100)
\end{lstlisting}
\normalsize
\end{tcolorbox}
\caption{Python TensorFlow Code: An example of a successful extraction of a training sample from the backdoored code generation LLM (perfect reconstruction).}
\label{fig:example2_tensorflow}
\end{figure}

\end{document}

%% file: training_alg.tex
\begin{algorithm}
	\caption{Model Training for a Memory Backdoor} \label{alg:training}
	\begin{algorithmic}[1]
            \State \textcolor{red!70!black}{$\mathcal{D}_t \gets \{ (x,y) \in \mathcal{D} \mid \text{criteria}(x,y) \}$} \Comment{select samples}
            \State \textcolor{red!70!black}{$\mathcal{I} \leftarrow \text{build}(\mathcal{D}_t)$} \Comment{build index}
		\For {$\text{epoch}=1, 2, \ldots$} 
			\For {$(\mathbf{X}_{\text{batch}}, \mathbf{Y}_{\text{batch}}) \in \mathcal{B}(\mathcal{D})$} \Comment{iterate over batches}
				\State $\mathbf{Y}'_{\text{batch}} \leftarrow f_{\theta}(\mathbf{X}_{\text{batch}})$
                \State $\mathcal{L}_{\text{total}} \leftarrow \mathcal{L}_{CE}(\mathbf{Y}'_{\text{batch}}, \mathbf{Y}_{\text{batch}})$
                \State \textcolor{red!70!black}{$\mathcal{I}_{\text{batch}} \leftarrow \mathcal{B}(\mathcal{I})$}\Comment{batch of indexes}
\State \textcolor{red!70!black}{$\mathcal{L}_{\text{mem}} \leftarrow \sum_{\iota \in \mathcal{I}_{\text{batch}}} \left[\mathcal{L}_1(f_\theta(G(\iota)),p_\iota) \right.$} \\
\textcolor{red!70!black}{$\hspace{9.5em} + \left. \mathcal{L}_2(f_\theta(G(\iota)),p_\iota)\right]$} 
                \State \textcolor{red!70!black}{$\mathcal{L}_{\text{total}} \leftarrow \mathcal{L}_{\text{total}} + \lambda \cdot \mathcal{L}_{\text{mem}}$}

                \State $\theta \leftarrow \text{Optimize}(\mathcal{L}_{\text{total}})$
			\EndFor
		\EndFor
	\end{algorithmic} 
	
\end{algorithm}

%% file: tabels/attack_tab.tex
\begin{table}[t]
\caption{The performance of the classification and segmentation vision models before and after a 1000 image memory backdoor attack (single client).}
\vspace{-.3em}
\label{tab:attack}
\resizebox{\columnwidth}{!}{%
\setlength{\tabcolsep}{3pt} 

\begin{tabular}{ccccccc}
               & \multicolumn{1}{c|}{}                 & \multicolumn{3}{c|}{Primary Task}                                           & \multicolumn{2}{c|}{Backdoor Task}                \\
               & \multicolumn{1}{c|}{}                 & \multicolumn{3}{c|}{\textit{ACC}}                                           & \textit{SSIM} & \multicolumn{1}{c|}{\textit{MSE}} \\ \hline\hline
\textbf{Dataset} & \multicolumn{1}{c|}{\textbf{Model}} & \textbf{Clean} & \textbf{Backdoored} & \multicolumn{1}{c|}{\textbf{Delta}}  & \multicolumn{2}{c|}{\textbf{Backdoored}}          \\ \hline\hline
MNIST          & \multicolumn{1}{c|}{CNN}              & 0.992          & 0.989              & \multicolumn{1}{c|}{\textbf{-0.003}} & 0.918         & \multicolumn{1}{c|}{0.011}           \\
\textbf{}      & \multicolumn{1}{c|}{FCN}              & 0.984          & 0.977              & \multicolumn{1}{c|}{\textbf{-0.007}} & 0.968        & \multicolumn{1}{c|}{0.003}            \\ \hline
CIFAR100       & \multicolumn{1}{c|}{CNN}              & 0.611          & 0.619               & \multicolumn{1}{c|}{\textbf{0.008}}   & 0.541         & \multicolumn{1}{c|}{0.011}            \\
\textbf{}      & \multicolumn{1}{c|}{VGG16}            & 0.652          & 0.615               & \multicolumn{1}{c|}{\textbf{-0.037}}  & 0.384         & \multicolumn{1}{c|}{0.040}            \\
\textbf{}      & \multicolumn{1}{c|}{VIT}              & 0.714          & 0.642               & \multicolumn{1}{c|}{\textbf{-0.072}}  & 0.991         & \multicolumn{1}{c|}{0.000}            \\ \hline
VGG FACE       & \multicolumn{1}{c|}{VIT}              &0.7             & 0.632              & \multicolumn{1}{c|}{\textbf{-0.068}}  & 0.853         & \multicolumn{1}{c|}{0.002}            \\ \hline
               &                                       &                &                     &                                      &               &                                   \\
               & \multicolumn{1}{c|}{}                 & \multicolumn{3}{c|}{\textit{DICE}}                                                   & \textit{SSIM} & \multicolumn{1}{c|}{\textit{MSE}} \\ \hline\hline
MRI            & \multicolumn{1}{c|}{VIT-S}            & 0.877          & 0.856             & \multicolumn{1}{c|}{\textbf{-0.021}} & 0.911       & \multicolumn{1}{c|}{0.001}           \\ \hline\hline
\end{tabular}%
}
\vspace{-1em}
\end{table}

%% file: tabels/ablation_study/song_comparison.tex


\begin{table}[t]
\centering
\caption{Comparison of task accuracy and reconstruction quality (denoted ACC/SSIM) across the baselines without pruning. Bold values indicate the best result in the experiment.}
\label{tab:song_whitebox_comparison}
\resizebox{\columnwidth}{!}{%
\begin{tabular}{lcccc}
\toprule
\textbf{No Pruning} & \textbf{LSB} & \textbf{Corr} & \textbf{Sign} & \textbf{Ours} \\
\midrule
CIFAR100 -- ViT & 66.20 / \textbf{1.00} & 65.21 / 0.7523 & 64.67 / 0.9757 & \textbf{67.32} / 0.9984 \\
MNIST -- FCN    & 97.89 / \textbf{1.00} & 98.00 / 0.9853 & 97.96 / 0.9892 & \textbf{98.13} / 0.9989 \\
MRI -- ViT      & \textbf{87.11} / \textbf{1.00} & 86.90 / 0.5872 & 86.75 / 0.8223$^{*}$ & 85.39 / 0.9931 \\
\midrule
\textbf{With Pruning} & \textbf{LSB} & \textbf{Corr} & \textbf{Sign} & \textbf{Ours} \\
\midrule
CIFAR100 -- ViT & 65.75 / 0.5645 & 64.95 / 0.6665 & 64.68 / 0.5702 & \textbf{66.78} / \textbf{0.7355} \\
MNIST -- FCN    & 97.88 / 0.7123 & 98.02 / 0.5067 & 98.08 / 0.5385 & \textbf{98.17} / \textbf{0.9975} \\
MRI -- ViT      & \textbf{87.14} / 0.5967 & 86.88 / 0.5786 & 86.72 / 0.5101$^{*}$ & 85.09 / \textbf{0.7172} \\
\bottomrule
\end{tabular}
}
\\[2mm]
\parbox{0.9\columnwidth}{\footnotesize *:\textit{ 10 images were used instead of 100 because the model weights cannot store more image bits.}}
\vspace{-1em}
\end{table}

%% file: tabels/ablation_study/DPSGD.tex
\begin{table}[t]
    \centering
    \caption{MNIST-FCN backdoor performance with/without DP-SGD, 
    for 0, 6{,}000 and 12{,}000 memorized samples.}
    \label{tab:mnist_fcn_dpsgd}

   \resizebox{0.6\columnwidth}{!}{%

    \small                          
    \setlength{\tabcolsep}{3pt}
    \renewcommand{\arraystretch}{0.85} %

    \begin{tabular}{@{}ccccc@{}}
        \toprule
        \# memorized & Attack & DPSGD & ACC & SSIM \\
        \midrule
        \multirow{2}{*}{0}
          & \xmark & \xmark & 0.985  & --    \\
          & \xmark & \cmark & 0.943 & --    \\
        \midrule
        \multirow{2}{*}{6,000}
          & \cmark & \xmark & 0.981 & 0.834 \\
          & \cmark & \cmark & 0.905   & 0.628 \\
        \midrule
        \multirow{2}{*}{12,000}
          & \cmark & \xmark & 0.981  & 0.725 \\
          & \cmark & \cmark & 0.903 & 0.637 \\
        \bottomrule
    \end{tabular}}
    \vspace{-1em}
\end{table}

%% file: tabels/covert_overt.tex
\begin{table}[h]
\setlength\tabcolsep{2 pt}
\caption{A performance comparison between using a visual index-trigger (Ours) and using an index code (TA from \cite{amittranspose}) as the trigger. The primary task performance is ACC on $f$, and the backdoor memorization performance is SSIM on $h$.}
\label{tab:covert_overt}
\centering
\resizebox{0.85\columnwidth}{!}{%
\begin{tabular}{ccccccc}
 &  & \multicolumn{1}{c|}{} & \multicolumn{2}{c|}{\textbf{$f$ ACC}} & \multicolumn{2}{c|}{\textbf{$h$ SSIM}} \\
Dataset & Model & \multicolumn{1}{c|}{$|\mathcal{D}_t|$} & Ours & \multicolumn{1}{c|}{TA} & Ours & \multicolumn{1}{c|}{TA} \\ \hline\hline
\multirow{4}{*}{CIFAR-100} & CNN & \multicolumn{1}{c|}{100} & \textbf{0.615} & \multicolumn{1}{c|}{0.572} & \textbf{0.827} & \multicolumn{1}{c|}{0.484} \\ \cline{2-7} 
 & VGG & \multicolumn{1}{c|}{200} & \textbf{0.633} & \multicolumn{1}{c|}{0.385} & \textbf{0.719} & \multicolumn{1}{c|}{0.202} \\ \cline{2-7} 
 & \multirow{2}{*}{VIT} & \multicolumn{1}{c|}{2000} & \textbf{0.619} & \multicolumn{1}{c|}{0.613} & \textbf{0.977} & \multicolumn{1}{c|}{0.685} \\
 &  & \multicolumn{1}{c|}{5000} & \textbf{0.622} & \multicolumn{1}{c|}{0.605} & \textbf{0.915} & \multicolumn{1}{c|}{0.592} \\ \hline\hline
\end{tabular}%
}
\end{table}
\setlength{\textfloatsep}{10pt}

%% file: tabels/ablation_study/purning.tex
\begin{table}[h]
  \centering
  \small
  \caption{Effect of global L1 pruning on attack performance}
  \label{tab:prune-combined-delta}
  \begin{minipage}{0.48\textwidth}
    \centering
    \textbf{(a) CIFAR‑100 ViT (1000 samples)}\\[0.5em]
    \begin{tabular}{ccccccc}
      \toprule
      Sparsity & ACC   & \(\Delta\)ACC & SSIM   & \(\Delta\)SSIM & MSE     & \(\Delta\)MSE \\
      \midrule
      0\%   & 0.673 & 0.000  & 0.998 & 0.000   & 0.000 & 0.000 \\
      5\%   & 0.673 & +0.000 & 0.991 & –0.007  & 0.000 & +0.000 \\
     10\%   & 0.673 & +0.001 & 0.926 & –0.072  & 0.000 & +0.000 \\
     15\%   & 0.671 & –0.002 & 0.817 & –0.181  & 0.001 & +0.001 \\
     20\%   & 0.669 & –0.004 & 0.736 & –0.263  & 0.002 & +0.002 \\
     25\%   & 0.668 & –0.005 & 0.619 & –0.380  & 0.005 & +0.005 \\
      \bottomrule
    \end{tabular}
  \end{minipage}\hfill
  \begin{minipage}{0.48\textwidth}
    \centering \vspace{.5em}
    \textbf{(b) MNIST FCN (3000 samples)}\\[0.5em]
    \begin{tabular}{ccccccc}
      \toprule
      Sparsity & ACC   & \(\Delta\)ACC & SSIM   & \(\Delta\)SSIM & MSE    & \(\Delta\)MSE \\
      \midrule
      0\%   & 0.981 & 0.000  & 0.725 & 0.000   & 0.045 & 0.000 \\
      5\%   & 0.981 & –0.000 & 0.725 & –0.000  & 0.045 & +0.000 \\
     10\%   & 0.981 & –0.000 & 0.724 & –0.000  & 0.045 & +0.000 \\
     15\%   & 0.981 &  0.000 & 0.724 & –0.000  & 0.045 & +0.000 \\
     20\%   & 0.981 & +0.000 & 0.722 & –0.002  & 0.045 & +0.000 \\
     25\%   & 0.982 & +0.001 & 0.722 & –0.003  & 0.045 & +0.000 \\
      \bottomrule
    \end{tabular}
  \end{minipage}
\end{table}